\documentclass[10pt,twocolumn,letterpaper]{article}

\usepackage{authblk}

\usepackage[pagenumbers]{iccv}          

\usepackage{multirow}
\usepackage{color}
\usepackage{makecell}

\usepackage{booktabs}

\definecolor{iccvblue}{rgb}{0.21,0.49,0.74}
\usepackage[pagebackref,breaklinks,colorlinks,allcolors=iccvblue]{hyperref}

\title{Lightweight and Fast Real-time Image Enhancement via Decomposition of the Spatial-aware Lookup Tables}

\author[1, 3]{Wontae Kim}
\author[2]{Keuntek Lee}
\author[1, 2]{Nam Ik Cho}
 
\affil[1]{IPAI, Seoul National University, Seoul, Korea} 
\affil[2]{Department of ECE, INMC, Seoul National University, Seoul, Korea}
\affil[3]{LG Electronics, Seoul, Korea \protect\\{\tt\small \{munte2,leekt000,nicho\}@snu.ac.kr}}

\begin{document}
\maketitle
\begin{abstract}
The image enhancement methods based on 3D lookup tables (3D LUTs) efficiently reduce both model size and runtime by interpolating pre-calculated values at the vertices. 
However, the 3D LUT methods have a limitation due to their lack of spatial information, as they convert color values on a point-by-point basis. 
Although spatial-aware 3D LUT methods address this limitation, they introduce additional modules that require a substantial number of parameters, leading to increased runtime as image resolution increases. 
To address this issue, we propose a method for generating image-adaptive LUTs by focusing on the redundant parts of the tables. 
Our efficient framework decomposes a 3D LUT into a linear sum of low-dimensional LUTs and employs singular value decomposition (SVD). 
Furthermore, we enhance the modules for spatial feature fusion to be more cache-efficient. 
Extensive experimental results demonstrate that our model effectively decreases both the number of parameters and runtime while maintaining spatial awareness and performance. 
The code is available at \url{https://github.com/WontaeaeKim/SVDLUT}.
\end{abstract}

\section{Introduction}
\label{sec:intro}

Automatic photo enhancement has gained attention because it improves the aesthetic quality of images and minimizes the need for manual retouching. 
Deep learning-based enhancement methods \cite{akyuz2020deep,a2021two,cai2018learning,chen2018deep,chen2018learning,deng2018aesthetic,
chen2021hdrunet,huang2022hdr,ignatov2017dslr,jiang2021enlightengan,kim2020jsi,
kim2020pienet,le2023single,liu2020single,moran2020deeplpf,wei2018deep,
yan2016automatic,zhang2019kindling} have shown impressive results.
Especially, 3D LUT methods \cite{liu20234d,liang2021ppr10k,kim2024image,wang2021real,yang2022adaint,yang2022seplut,yang2024taming,zeng2020lut,zhang2022clut} demonstrate an excellent balance between performance, model size, and inference time, even in tasks other than image enhancement \cite{Jo_2021_CVPR, li2022mulut,ma2022learning,liu2023reconstructed,Li_2024_CVPR}.
These methods enable the direct retrieval of output values through interpolation, eliminating the need for repeated, complex computations. 
This capability effectively replaces a non-linear color transformation, resulting in reduced computational costs and faster inference times.

\begin{figure}[tb]
  \centering
  \includegraphics[width=0.75\columnwidth]{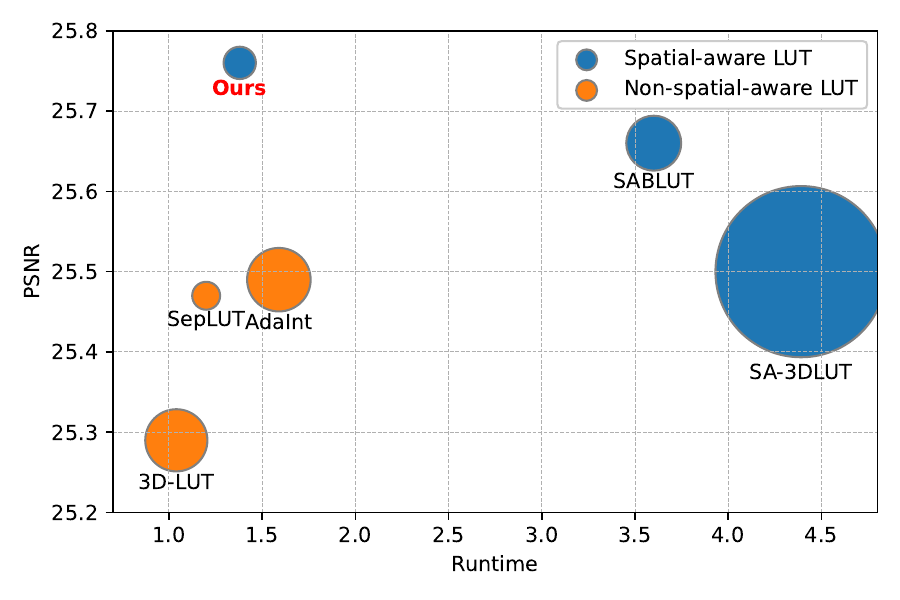}
  \caption{Trade-off between performance and runtime on FiveK dataset \cite{bychkovsky2011learning}. The circle size is proportional to the model size.}
  \label{fig:tradoff}
\end{figure}

Notably, the image-adaptive 3D LUT method \cite{zeng2020lut} outperforms fixed 3D LUT methods by generating a customized 3D LUT for each individual image. 
However, this method has a limitation in that it does not utilize spatial information. 
To address this limitation, Wang \etal investigated a spatial-aware 3D LUT method \cite{wang2021real} that generates and fuses spatial features before applying the 3D LUT for image enhancement. 
However, the additional processes required for spatial awareness lead to an increase in the number of parameters and inference time, representing a trade-off. 
More recently, Kim \etal \cite{kim2024image} have effectively reduced model size; however, the method still suffers from long inference times when processing high-resolution images.
To address the limitation of spatial-aware LUT methods, we present a framework that features a compact model size and short inference time while maintaining competitive performance, as illustrated in \cref{fig:tradoff}.
To enhance efficiency, we analyze previous LUT-based image enhancement methods and identify two key insights. 
First, the 3D LUTs generated for each image can often be redundant and can be effectively replaced with linear operations involving 2D LUTs. 
Moreover, inspired by singular value decomposition (SVD), the 2D LUT can be substituted with a diagonal matrix and two sparse matrices, resulting in an 88\% reduction in the number of parameters compared to the 3D LUT. 
Second, we observe that previous methods are not cache-efficient when incorporating spatial information, which significantly contributes to long inference times, especially for high-resolution images. 
Therefore, we propose a cache-efficient spatial information fusion structure that delivers quick inference, even for high-resolution images.

The contributions of this paper are three-fold:
\begin{itemize}
\item We introduce a new approach that replaces a 3D LUT with 2D LUTs, which are further simplified by SVD.
\item We propose an efficient architecture for the LUT-based image enhancement method that achieves spatial awareness while maintaining a compact model size and short inference time.
\item We conduct extensive experiments to evaluate our approach against existing methods using two benchmark datasets. The results demonstrate the efficiency and effectiveness of our method in image enhancement.
\end{itemize}

\begin{figure*}[tb]
    \centering   

    \begin{minipage}{0.7\linewidth}
        \begin{subfigure}{\linewidth}
            \includegraphics[width=\columnwidth]{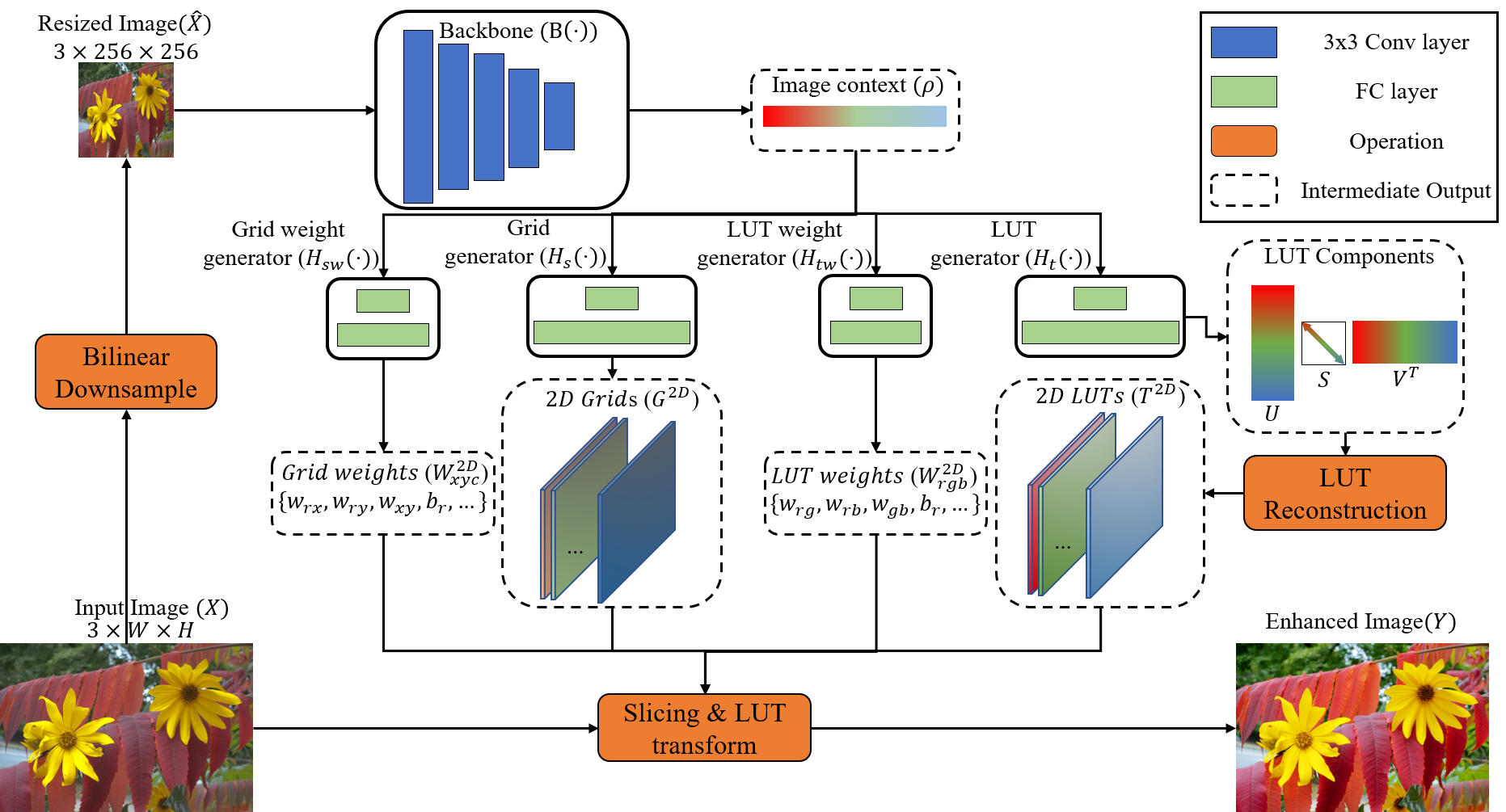}
            \caption{Overall structure}
            \label{fig:framework}
        \end{subfigure}
    \end{minipage}
    \hfill
    \begin{minipage}{0.28\linewidth}
        \begin{subfigure}{\linewidth}
            \includegraphics[width=\columnwidth]{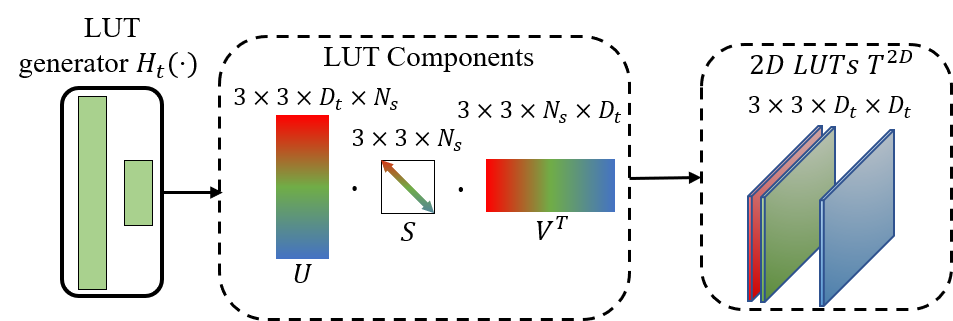}
            \caption{LUT generation with SVD}
            \label{fig:lutgen_svd}
        \end{subfigure}
        \vfill
        \begin{subfigure}{\linewidth}
            \includegraphics[width=\columnwidth]{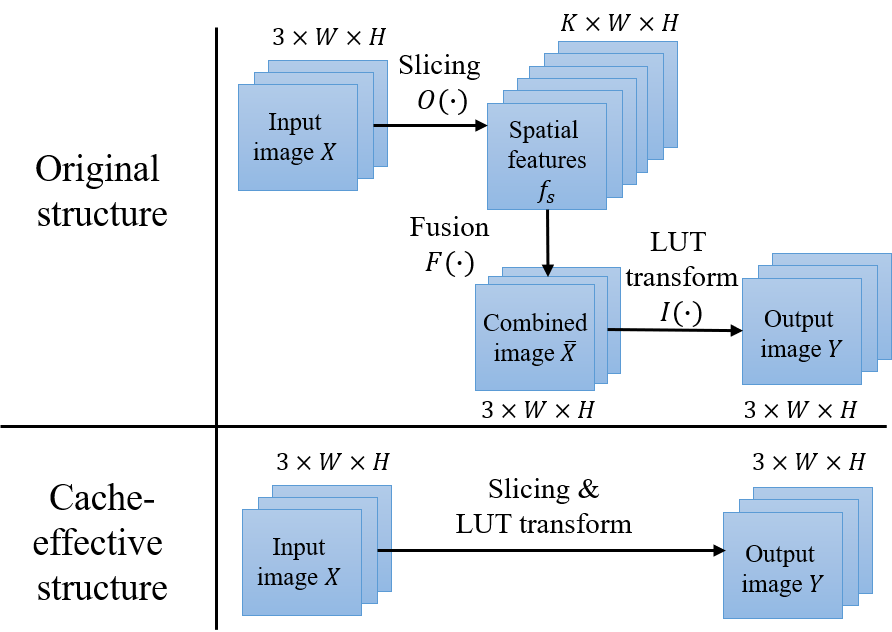}
            \caption{Cache-effective structure}
            \label{fig:cache}
        \end{subfigure}
    \end{minipage}

    \caption{(a) Overall structure of our SVDLUT. The backbone network extracts image context from resized images. 
    The image context is used to predict bilateral grids, LUT components, and weights of LUTs, and bilateral grids.
    LUTs are reconstructed from LUT components. The input image is enhanced by slicing and LUT transform operation.
    (b) Process of LUT generation using SVD.
    (c) Comparison of our cache-effective structure with the original.}
\end{figure*}

\section{Related Work}
\label{sec:realted_work}

\subsection{Image Enhancement Methods with 3D LUT}
Image enhancement methods based on 3D LUTs have gained attention due to their efficient operation and real-time performance \cite{liu20234d,liang2021ppr10k,kim2024image,wang2021real,
yang2022adaint,yang2022seplut,zeng2020lut,zhang2022clut}. An LUT-based method incorporating deep learning has also been introduced, which predicts weights by a simple CNN and generates image-adaptive 3D LUTs through a weighted sum of basis 3D LUTs \cite{zeng2020lut}.
Yang \etal enhanced the capacity of LUTs by combining 1D LUTs \cite{yang2022seplut} or adaptive vertices \cite{yang2022adaint} with 3D LUTs. 
While these advanced LUT methods significantly improve performance, they still face limitations due to a lack of spatial information. 

To overcome the limitation, Wang \etal proposed the spatial-aware 3D LUT method (SA-3DLUT) \cite{wang2021real}, which generates spatial features using a U-Net \cite{ronneberger2015u} style backbone and combines these features with the input image. 
Although the spatial awareness tools enhance performance, they require a significant number of parameters and result in longer inference times.
Meanwhile, Kim \etal replaced the U-Net style decoder with bilateral grids and a slicing operation \cite{kim2024image}. 
This approach effectively reduces the number of parameters; however, the inference time for high-resolution images remains longer than that of other non-spatial-aware methods. 
Given this trade-off, we investigate the increase in model size and inference time associated with spatial awareness. 
Based on these insights, we optimize the process of generating spatial awareness using a linear operation of 2D LUT combined with a cache-efficient structure.

\subsection{LUT Size Reduction}
\label{sec:size_reduction}
The parameters of 3D LUT-based models are mainly used to generate the values in the table, and reducing the table size is one of the most effective ways to decrease the model's overall size. 
A straightforward method for reducing the size of an LUT is to decrease the number of sampling points; however, this can lead to a decline in performance.
Zhang \etal \cite{zhang2022clut} demonstrated that the output color shows different correlations within each axis; specifically, a strong correlation exists within the same channel axis, while there is a weaker correlation with the other axes.
Building on this finding, they proposed the  CLUT~\cite{zhang2022clut}, which uses varying bin sizes based on correlation and decomposes the LUT into one basis LUT and two transformation matrices. 
Furthermore, Li \etal \cite{Li_2024_CVPR} effectively reduced the size of LUTs in super-resolution tasks by employing the Diagonal First Compression (DFC) framework.
They found that most pixel values correspond to the entries along the diagonal of the LUTs. 
Their approach involved compressing the LUTs by subsampling the non-diagonal values and re-indexing the remaining entries. 
While previous methods have focused on reducing the size of LUTs by subsampling axes with weak correlations, we propose a new strategy that replaces the 3D LUT with linear operations of 2D LUTs with their SVD.

\section{Proposed Method}
\subsection{Preliminary: Image-adaptive 3D LUT Methods}
\label{sec:pre}
Most 3D LUT methods \cite{kim2024image,liang2021ppr10k,wang2021real,yang2022adaint,yang2022seplut,zeng2020lut} enhance the image quality by the following three steps.
First, they extract image context feature $\rho$ from a resized input image $\hat{X}$, which can be expressed as
\begin{equation}
    \rho = B(\hat{X}),
    \label{eq:backbone}
\end{equation}
where $B(\cdot)$ is a simple backbone network. The backbone network can handle images of arbitrary resolution and maintains a small size by using resized images as input.

Second, preprocessing is performed, where various methods have been proposed to enhance its efficiency, such as spatial feature fusion \cite{wang2021real,kim2024image}, 1D LUT processing \cite{yang2022seplut}, or adaptive vertices \cite{yang2022adaint}.
Especially, spatial feature fusion is used to provide spatial information in spatial-aware 3D LUT methods, which combines an input image and spatial-aware maps $f_s \in \mathbb{R}^{K \times W \times H}$.
Spatial-aware maps can be generated by slicing with bilateral grids \cite{kim2024image} or U-net style decoder \cite{wang2021real}. 
This step can be described as 
\begin{equation}
    \bar{X} = F(X,f_s)   
    \label{eq:spatialfusion}
\end{equation}
where $F(\cdot)$ is preprocessing operation and $\bar{X}$ is a preprocessed image.

Finally, they generate image-adaptive 3D LUTs $T^{3D} \in \{t^{c}_{rgb}|c \in \{r, g, b\} \}$ from context features $\rho$, where $t^{c}_{rgb}$ is an LUT which has $r$-, $g$-, and $b$-axis for output channel $c$.
A 3D LUT $t^{c}_{rgb} \in \mathbb{R}^{D_{t} \times D_{t} \times D_{t}}$ is composed of sparsely sampled enhanced color values with $D_t$ vertices on each axis.
Each 3D LUT is used to retrieve the enhanced color values by interpolation. 
\begin{equation}
    Y_{(c,x,y)} = I_{tri}(\bar{X}_{(c,x,y)},t^{c}_{rgb}), \quad T^{3D} = H_{t}(\rho),  
    \label{eq:coloren}
\end{equation}
where $Y_{(c,x,y)}$ is the output color value at the corresponding coordinate $(c,x,y)$, and $I_{tri}(\cdot)$ is trilinear interpolation.

\subsection{SVDLUT}
\label{sec:SVDLUT}
The \cref{fig:framework} shows our spatial-aware 3D LUT framework. 
Our method effectively addresses the inherent limitations of spatial-aware LUT methods by reducing the number of parameters through decomposition methods and minimizing inference time at high resolution with a cache-effective structure.
First of all, image context $\rho$ is extracted by the same backbone network of SABLUT~\cite{kim2024image} as described in \cref{eq:backbone}, since this aspect is not our primary contribution.
Second, the generators create LUT components, bilateral grids, and their corresponding weights from the image context. 
In this operation, we analyze the utilization of 3D LUTs and suggest a decomposition method using low-dimensional LUTs in \cref{sec:decompose}.
Furthermore, the decomposition method by SVD and the corresponding analysis are also provided in \cref{sec:SVD}.
The bilateral grid with slicing is selected to provide spatial awareness, as its structural similarity to LUTs enables the use of the same decomposition techniques.
Finally, an input image is enhanced by slicing and LUT transform operation.
We found that previous spatial feature fusion incurs a long inference time due to poor cache effectiveness.
We introduce a cache-effective operation that combines the slicing and LUT transform in \cref{sec:cache_effective}.
The details of the network layers and hyperparameters are provided in the supplementary material.

\begin{table}[]
    \centering
    \caption{Ablation study under different dimensions of LUT and bilateral grid. Each cell indicates PSNR (model size) measured for the photo retouch task on the FiveK dataset.}
    \label{tab:abl_dimension}
    \resizebox{\columnwidth}{!}{%
    \begin{tabular}{c|c|c|c|c}
      \hline
      \multicolumn{2}{c|}{\multirow{2}{*}{}}& \multicolumn{3}{c}{Bilateral   grid}            \\ \cline{3-5}
      \multicolumn{2}{c|}{}                 & 3D             & 2D             & 1D             \\ \hline
      \multirow{3}{*}{LUT} & 3D & 25.68 (1.3M)   & 25.67 (1.1M)   & 25.54 (1.0M)   \\ \cline{2-5}
                           & 2D & 25.67 (421.5K) & 25.68 (205.3K) & 25.53 (161.2k) \\ \cline{2-5}
                           & 1D & 25.37 (335.9K) & 25.53 (119.8K) & 25.22 (75.7K)  \\ \hline
    \end{tabular}%
    }
\end{table}

\subsubsection{Decomposition into Lower Dimension}
\label{sec:decompose}
\begin{figure}[tb]
    \centering
    \begin{subfigure}{0.49\linewidth}
        \includegraphics[width=\columnwidth]{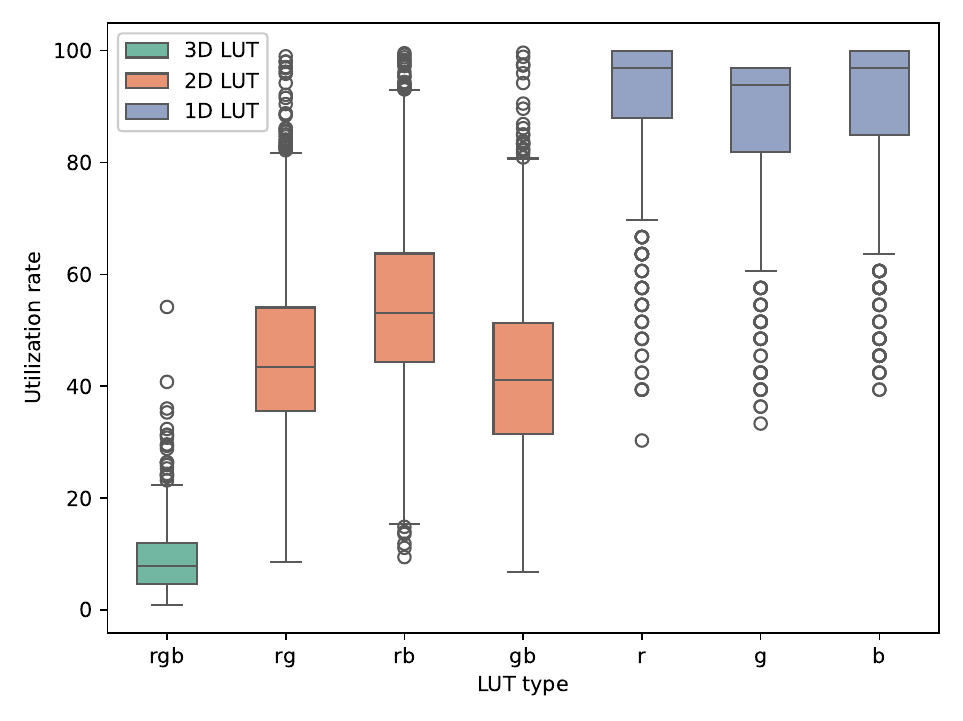}
        \caption{LUT utilization rate}
        \label{fig:lut_util}
    \end{subfigure}
    \hfill
    \begin{subfigure}{0.49\linewidth}
        \includegraphics[width=\columnwidth]{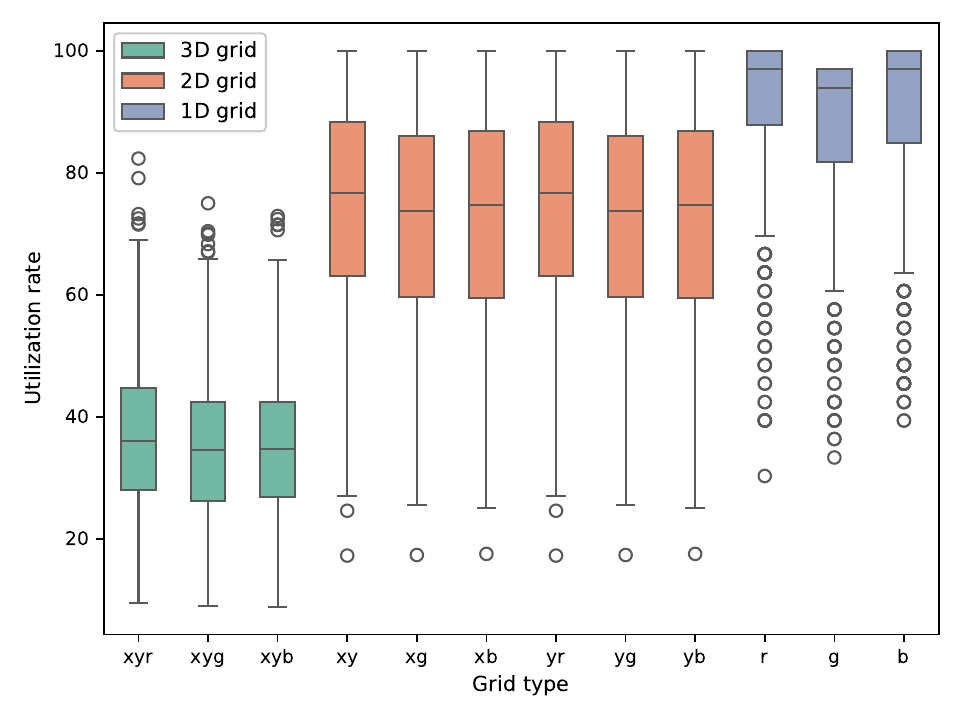}
        \caption{Grid utilization rate}
        \label{fig:grid_util}
    \end{subfigure}

    \begin{subfigure}{0.49\linewidth}
        \includegraphics[width=\columnwidth]{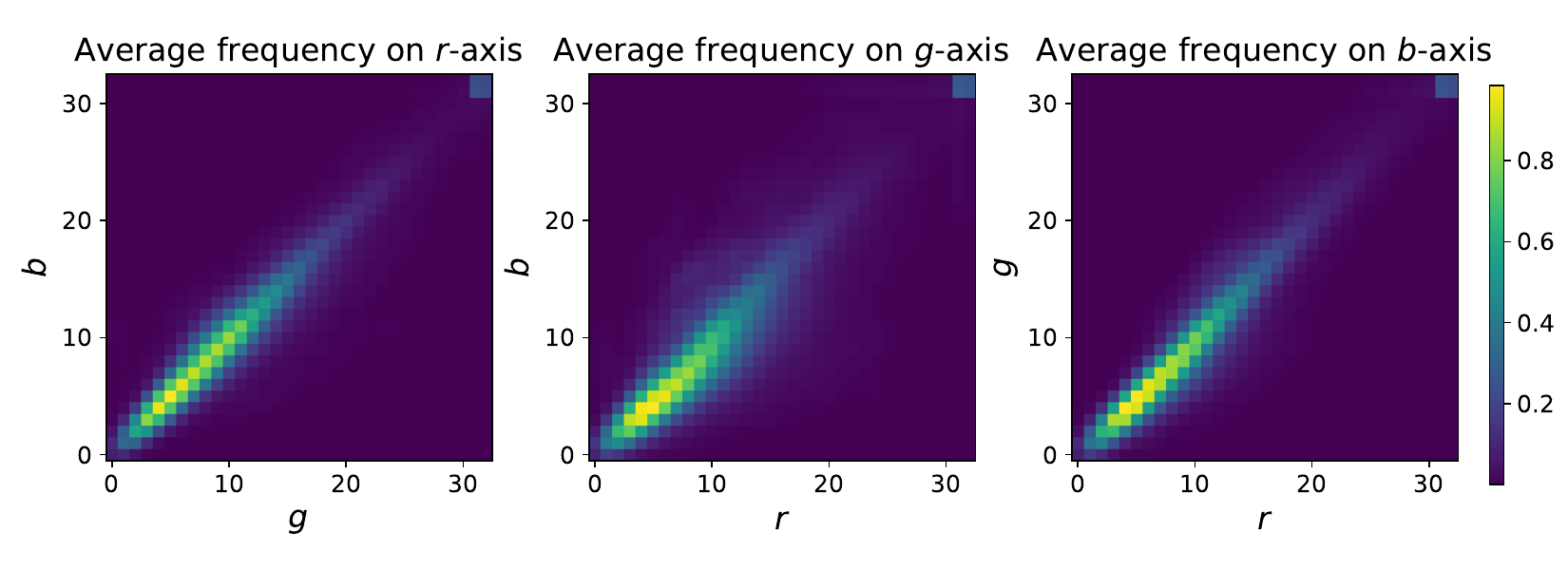}
        \caption{LUT occurrence statistics}
        \label{fig:lut_occurence}
    \end{subfigure}
    \hfill
    \begin{subfigure}{0.49\linewidth}
        \includegraphics[width=\columnwidth]{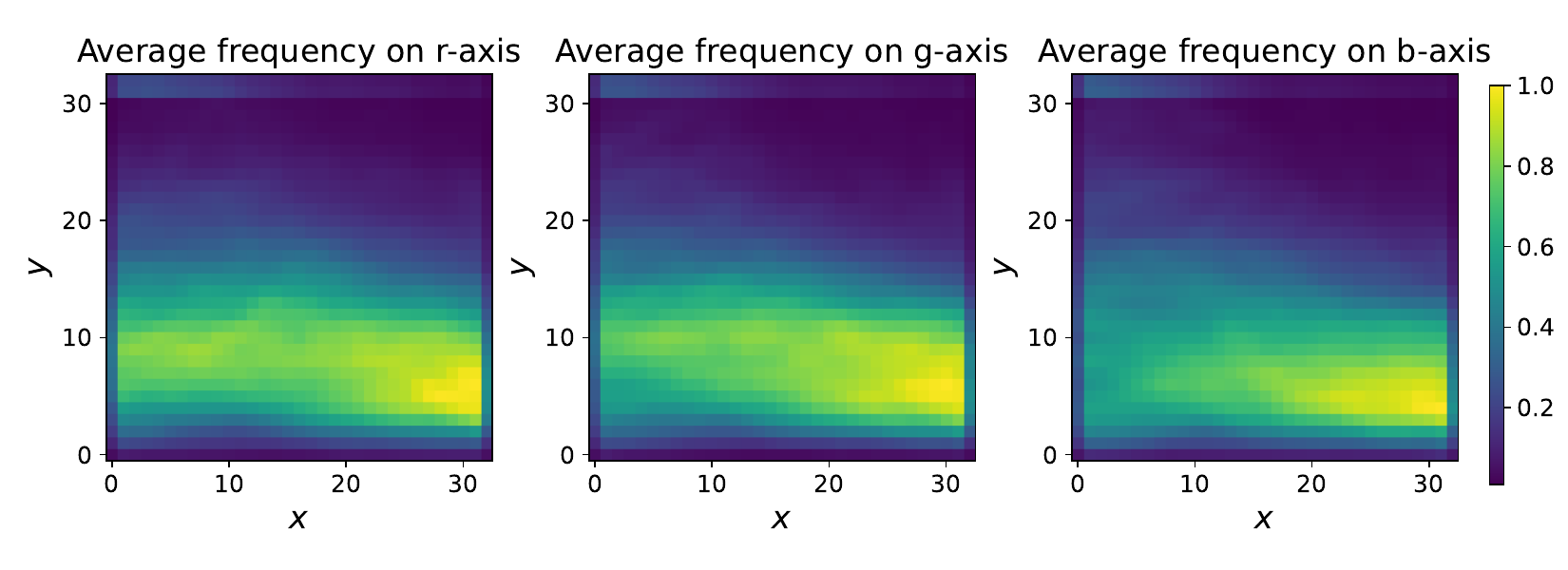}
        \caption{Grid occurrence statistics}
        \label{fig:grid_occurence}
    \end{subfigure}

    \caption{(a) and (b) are box plots of utilization rates on the FiveK dataset \cite{bychkovsky2011learning} under different dimensions of LUT and bilateral grid. 
            Each color represents the dimension of LUT and the bilateral grid. 
            (c) and (d) are LUT visualizations of average occurrence statistics on each axis for the FiveK.
            The cells closer to yellow indicate more frequently accessed vertices and the cells closer to blue indicate less frequently accessed vertices on predicted 3D LUTs.}
    \label{fig:util_rate}  
\end{figure}

For fixed LUT methods, 3D LUTs have been preferred to lower-dimensional LUTs due to their high capacity, which can universally handle multiple images \cite{vandenberg2018survey,yang2022seplut}. 
However, the size of the 3D LUTs significantly contributes to the overall model size and image-adaptive LUT is generated for a given image.
In this context, we analyze the utilization of the 3D LUT to identify ways to reduce the model size. 
We conduct an experiment to measure the utilization rate of the 3D LUT using the FiveK dataset \cite{bychkovsky2011learning}. 
The rate indicates the proportion of vertices in the 3D LUT that are actually referenced compared to the total number of generated vertices for each image $(\frac{\# referenced \ vertices}{\# generated \ vertices} \times 100)$.
As can be seen in \cref{fig:lut_util}, the vertices on a predicted 3D LUT are used only under 10\%.
Although 1D LUTs are fully utilized, it might imply that 1D LUTs are saturated and might not be enough to involve correlation among color channels. On the other hand, the 2D LUTs seem to have a proper utilization rate.

We also check the occurrence statistics that count the number of accesses for each vertex on 3D LUT and visualize it in \cref{fig:lut_occurence}.
Most of the higher occurrence frequencies are distributed along the diagonal. 
The results for PPR10K \cite{liang2021ppr10k} are similar to FiveK and are provided in supplementary material.
Two experiments in this section show that only a very small portion of the generated 3D LUT is used.
It suggests that the model size can be reduced by eliminating the wasted parameters to predict redundant 3D LUT vertices.

In this context, we formulate a hypothesis in which the combination of 2D LUTs $T^{2D} \in \{t^{c}_{rg},t^{c}_{rb},t^{c}_{gb}|c \in \{r,g,b\}\}$ can replace a 3D LUT like
\begin{equation}
    t^{c}_{rgb} \to w^{c}_{rg} \cdot t^{c}_{rg} + w^{c}_{rb} \cdot t^{c}_{rb} + w^{c}_{gb} \cdot t^{c}_{gb} + b^{c},
    \label{eq:LUTdecompose}
\end{equation} 
where $W^{2D}_{rgb} = \{w^{c}_{rg}, w^{c}_{rg},w^{c}_{gb},b^{c}|c \in \{r,g,b\}\}$ are weights and bias for the corresponding LUT.
Instead of \cref{eq:coloren}, the color enhancement for output channel $c$ by decomposed 2D LUT transform $Transform_{2D}^{c}(\bar{X}, T^{2D})$ can be formulated as
\begin{align}
    Y_{(c,x,y)} = & w^{c}_{rg} \cdot I_{bi}(\bar{X}_{(x,y)}, t^{c}_{rg})+ \nonumber \\
               & w^{c}_{rb} \cdot I_{bi}(\bar{X}_{(x,y)}, t^{c}_{rb})+ \nonumber \\
               & w^{c}_{gb} \cdot I_{bi}(\bar{X}_{(x,y)}, t^{c}_{gb})+b^{c},     
    \label{eq:LUTdecompose_out}
\end{align}
where $I_{bi}(\cdot)$ is a bilinear interpolation.

We also analyze bilateral grids $G^{3D} = \{g^{c'_k}_{xyc'_k}| 1 \le k \le K\}$ since they have similar structure \cite{chen2007real} to 3D LUTs and are effective ways to convey the spatial-awareness \cite{kim2024image}.
Although bilateral grids have a wider distribution than 3D LUT, as can be seen in \cref{fig:grid_util} and \cref{fig:grid_occurence}, the tendency is similar, \ie 3D bilateral grids are redundant and 1D bilateral grids are saturated.
We try to test bilateral grid decomposition to generate spatial features.
\cref{eq:spatialfusion} can be replaced with $Slicing_{2D}^{c'_k}(X, G^{2D})$  

\begin{figure}[tb]
    \centering
    \begin{subfigure}{0.48\linewidth}
        \includegraphics[width=\columnwidth]{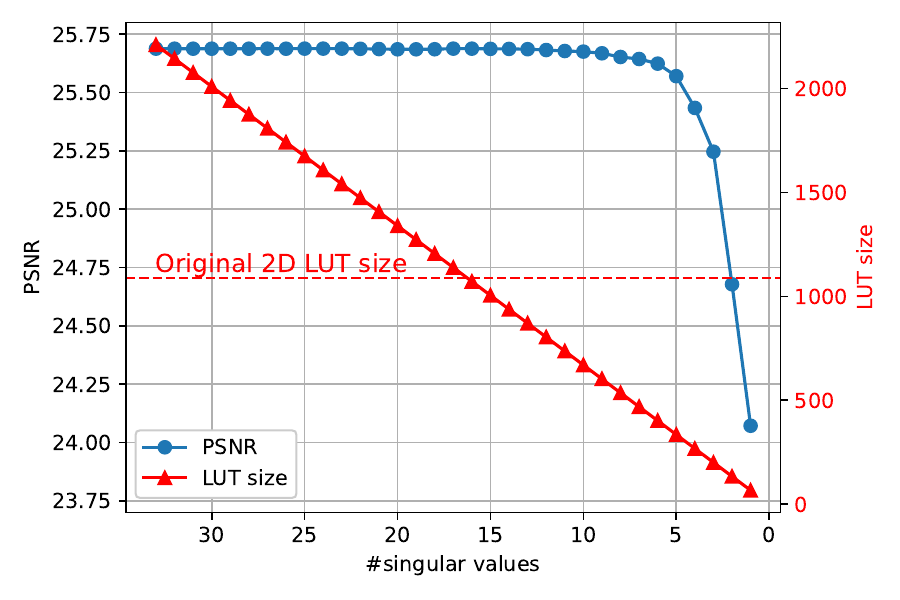}
        \caption{Toy experiment for LUT}
        \label{fig:lut_svd}
    \end{subfigure}
    \hfill
    \begin{subfigure}{0.48\linewidth}
        \includegraphics[width=\columnwidth]{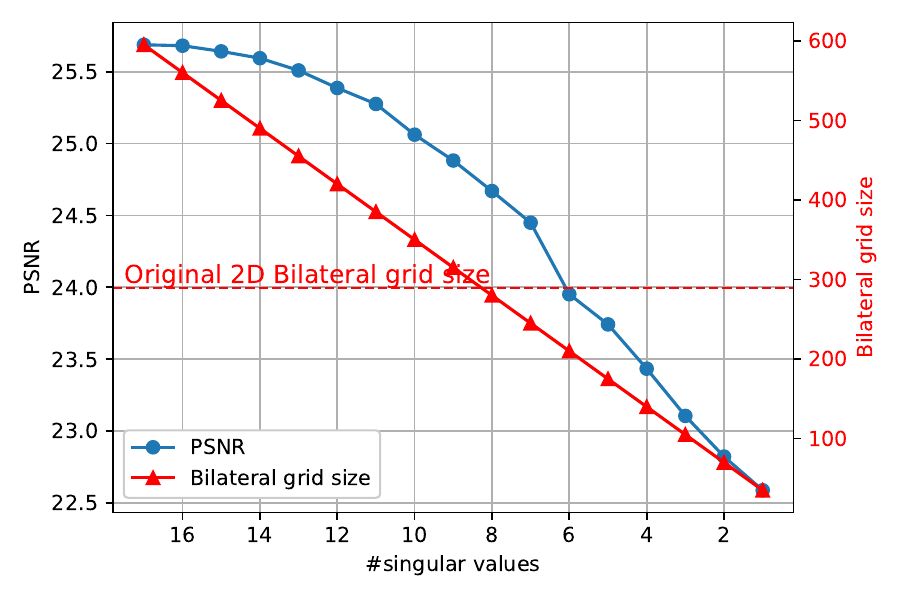}
        \caption{Toy experiment for Grid}
        \label{fig:grid_svd}
    \end{subfigure}

    \begin{subfigure}{0.49\linewidth}
        \includegraphics[width=\columnwidth]{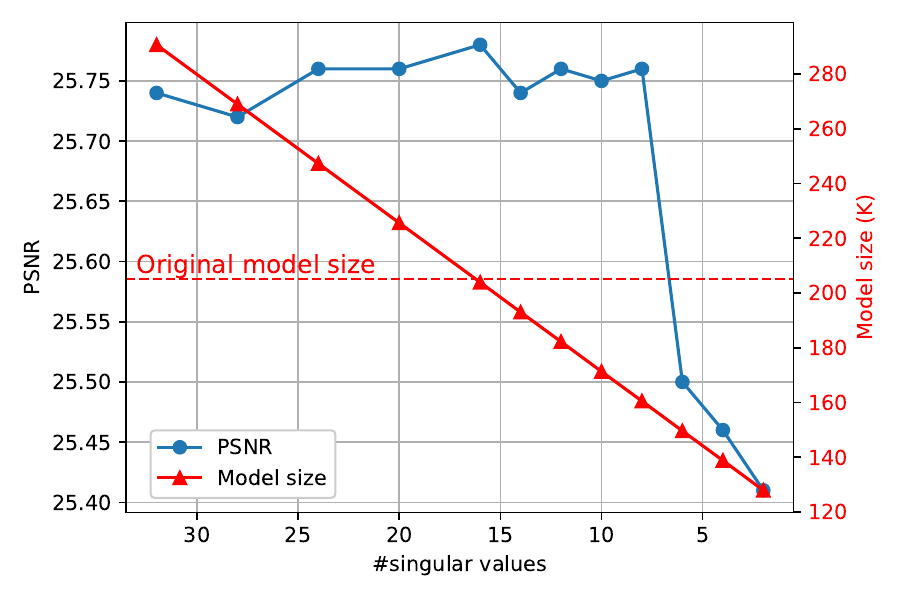}
        \caption{Trained model for LUT}
        \label{fig:abl_lut_singular}
    \end{subfigure}
    \hfill
    \begin{subfigure}{0.49\linewidth}
        \includegraphics[width=\columnwidth]{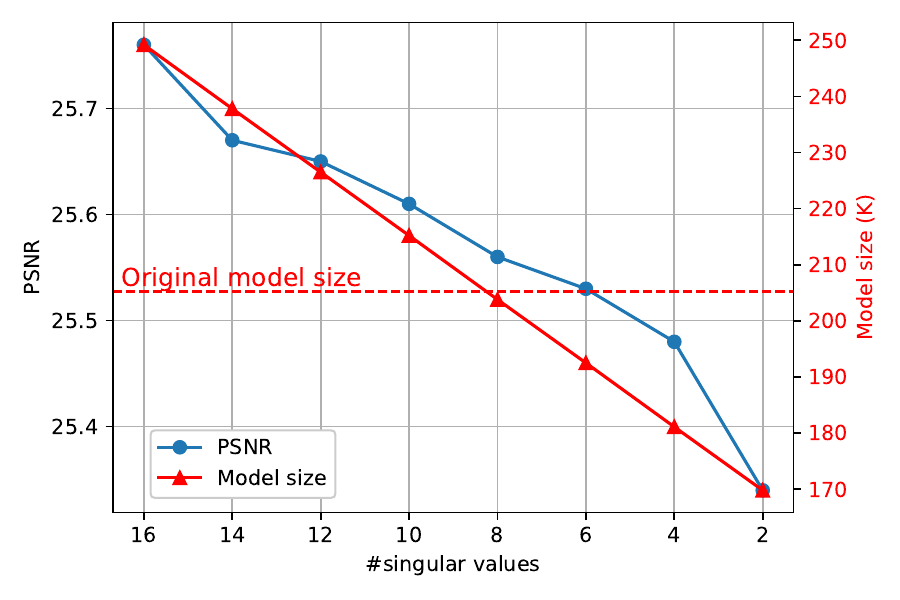}
        \caption{Trained model for Grid}
        \label{fig:abl_grid_singular}
    \end{subfigure}

    \caption{The results of the experiment to measure performance by reducing the singular value.
            (a) and (b) are the results of toy experiments for LUT and bilateral grid.
            (c) and (d) are the results of trained model for LUT and bilateral grid.
            Each blue line indicates PSNR with respect to singular value. Each red line represents the size of LUT, bilateral grid, and model size, respectively}
    \label{fig:svd_toy} 
\end{figure}

\begin{align}  
    g^{c'_k}_{xyc'_k} \to  w^{c'_k}_{xy} \cdot &g^{c'_k}_{xy} + w^{c'_k}_{xc'_k} \cdot g^{c'_k}_{xc'_k} + w^{c'_k}_{yc'_k} \cdot g^{c'_k}_{yc'_k} + b^{c'_k}, \label{eq:Griddecompose} \\
    f_{s(c'_k,x,y)}= & w^{c'_k}_{xy} \cdot I_{bi}(X_{(c'_{k'},x,y)}, g^{c'_k}_{xy}) + \nonumber \\ 
                    & w^{c'_k}_{xc'_k} \cdot I_{bi}(X_{(c'_{k'},x,y)}, g^{c'_k}_{xc'_k}) + \nonumber \\ 
                    &w^{c'_k}_{yc'_k} \cdot I_{bi}(X_{(c'_{k'},x,y)}, g^{c'_k}_{yc'_k})+b^{c'_k}, \label{eq:slicing} \\     
    \bar{X} = &Conv_{(K+3) \to 3}^{1 \times 1}(Concat(X,f_s)) \label{eq:fusion},
\end{align}
where $k'=mod(k,3)$, and $ Conv_{(K+3) \to 3}^{1 \times 1}(\cdot)$ is the 1$\times$1 convolution which has $(K + 3)$ input and 3 output channels.

We verify our hypothesis through following experiments.
We use the SABLUT \cite{kim2024image} as the baseline since this model utilizes the 3D LUT and 3D bilateral grid.
We assess the PSNR and model size under different dimensions of LUT and bilateral grid on FiveK photo retouch task.
As can be seen in \cref{tab:abl_dimension}, there is no significant change in PSNR when changing from 3D to 2D for both LUT and bilateral grid.
On the other hand, PSNR is degraded in the case of 1D LUT and bilateral grid.
We decide to use 2D for LUT and bilateral grid because this combination can reduce the model size by 84\% without performance degradation.

\subsubsection{Singular Value Decomposition (SVD)}
\label{sec:SVD}
After we decompose 3D LUTs into 2D LUTs and bilateral grids, we consider using the SVD to reduce the number of parameters.
To verify the efficiency of this approach, we conduct the following toy experiments.
We decompose the pre-trained 2D LUTs into singular values $S$ and two matrices $U$, $V$.
After reducing the singular values and corresponding rows of two matrices, we reconstruct the 2D LUT.
As can be seen in \cref{fig:lut_svd}, we measure the PSNR of the enhanced image by reconstructed 2D LUT.
\cref{fig:grid_svd} presents the results of the same experiment that applies SVD to the bilateral grid.
While the LUT maintains its performance with up to eight singular values, the bilateral grid experiences a performance drop from the beginning.

\cref{fig:abl_lut_singular} and \cref{fig:abl_grid_singular} illustrate the performance of models trained from scratch.
Specifically, the models predicts LUT components $(S,U,V)$ and reconstruct 2D LUTs from those components as illustrated in \cref{fig:lutgen_svd} like 
\begin{equation}
 T^{2D} = U \cdot S \cdot V^{T}, \quad S, U, V^{T} = H_t(\rho), 
 \label{eq:gen_LUT_SVD}
\end{equation}
where $V^T$ is transposed matrix of $V$.
These results have a similar tendency to toy examples.
The reason seems to be that the LUT has a monotonic and simple structure, whereas the bilateral grid incorporates spatial information.
The SVD decomposition is effective when singular values are smaller than 16 for LUT and 8 for bilateral grids.
Hence, we decide to apply SVD to the LUT but not to the bilateral grid.

\subsubsection{Cache-Effective Spatial Feature Fusion}
\label{sec:cache_effective}
In this section, we explore the processes involved in spatial-aware LUTs and enhance the structure to reduce runtime.
To conduct our investigation, we measure the runtime of various modules by adding them incrementally and compare our method with the original structure in \cite{kim2024image,wang2021real}, as shown in \cref{fig:cache}.
As illustrated in the left side of \cref{tab:runtime}, the runtime of the original structure tends to increase at higher resolutions. This increase is attributed to spatial feature fusion, slicing, and LUT transformations.
Specifically, these processes read high-resolution input and write high-resolution intermediate outputs ($f_s$ and $\bar{X}$), which result in frequent data exchanges between fast, high-rank memory and slow, low-rank memory within memory hierarchies. 
This frequent exchange is the main reason for the increase in runtime at high resolutions. 
To address this issue, we combine slicing and LUT transformation to reduce the runtime by minimizing the generation of intermediate outputs and reusing previously calculated results, such as indices of LUT.
Additionally, we remove the convolution in \cref{eq:fusion} since weighted sum for decomposition can not only replace it but also enhance image-adaptiveness.
We effectively reduce the runtime, as can be seen in the right side of \cref{tab:runtime}, and improve the PSNR by 0.08 dB for the photo retouch task on the FiveK dataset.
The cache-effective spatial-aware LUT transform can be described as
\begin{align}
    Y_{(c,x,y)} = Transform_{2D}^{c}(X_{(c,x,y)},T^{2D}) +  \nonumber \\
    \sum_{k=0}^{K/3-1} Slicing_{2D}^{c'_{c+3k}}(X_{(c,x,y)},G^{2D}). 
    \label{eq:slcing_n_lut_transform}
\end{align}
More details are provided in supplementary material.

\begin{table}[]
    \caption{Runtime across modules for original and cache-effective structure.}
    \label{tab:runtime}
    \resizebox{\columnwidth}{!}{%
    \begin{tabular}{c|c|c||c|c|c}
    \hline
    \multicolumn{3}{c||}{Original structure}               & \multicolumn{3}{c}{Cache-effective structure}                                              \\ \hline
    \multirow{2}{*}{Component}    & \multicolumn{2}{c||}{Runtime(ms)} & \multirow{2}{*}{Component}        & \multicolumn{2}{c}{Runtime(ms)}               \\ \cline{2-3} \cline{5-6}
                         & 480p           & 4K             &                                              & 480p                  & 4K                    \\ \hline \hline
    Backbone             & 0.72           & 0.73           & Backbone                                     & 0.72                  & 0.72                  \\
    Grid/weight   Gen    & 0.23           & 0.23           & Grid/weight   Gen                            & 0.23                  & 0.23                  \\
    Slicing              & 0.19           & 0.54           & LUT/weight   Gen                             & 0.3                   & 0.31                  \\
    Fusion               & 0.05           & 1.6            & \multirow{3}{*}{\makecell{Slicing \& \\ LUT Transform}} & \multirow{3}{*}{0.12} & \multirow{3}{*}{0.13} \\
    LUT/weight   Gen     & 0.29           & 0.05           &                                              &                       &                       \\
    LUT   Transform      & 0.09           & 0.7            &                                              &                       &                       \\ \hline
    Total                & 1.57           & 3.84           & Total                                        & 1.37                  & 1.38                  \\ \hline
    \end{tabular}%
    }
\end{table}

\begin{figure*}[tb]
    \centering
    \begin{subfigure}{0.49\linewidth}
        \includegraphics[width=\columnwidth]{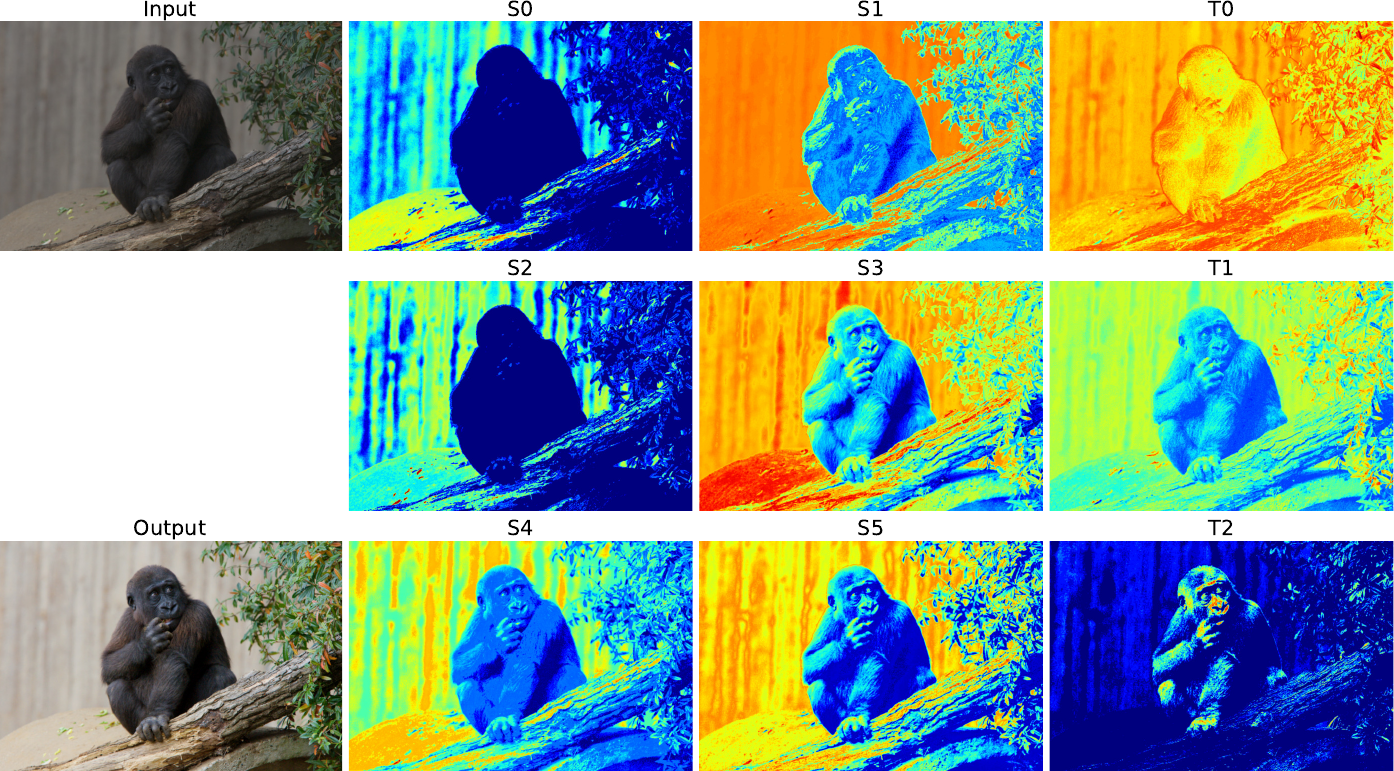}
        \caption{}
        \label{fig:analysis1}
    \end{subfigure}
    \hfill
    \begin{subfigure}{0.49\linewidth}
        \includegraphics[width=\columnwidth]{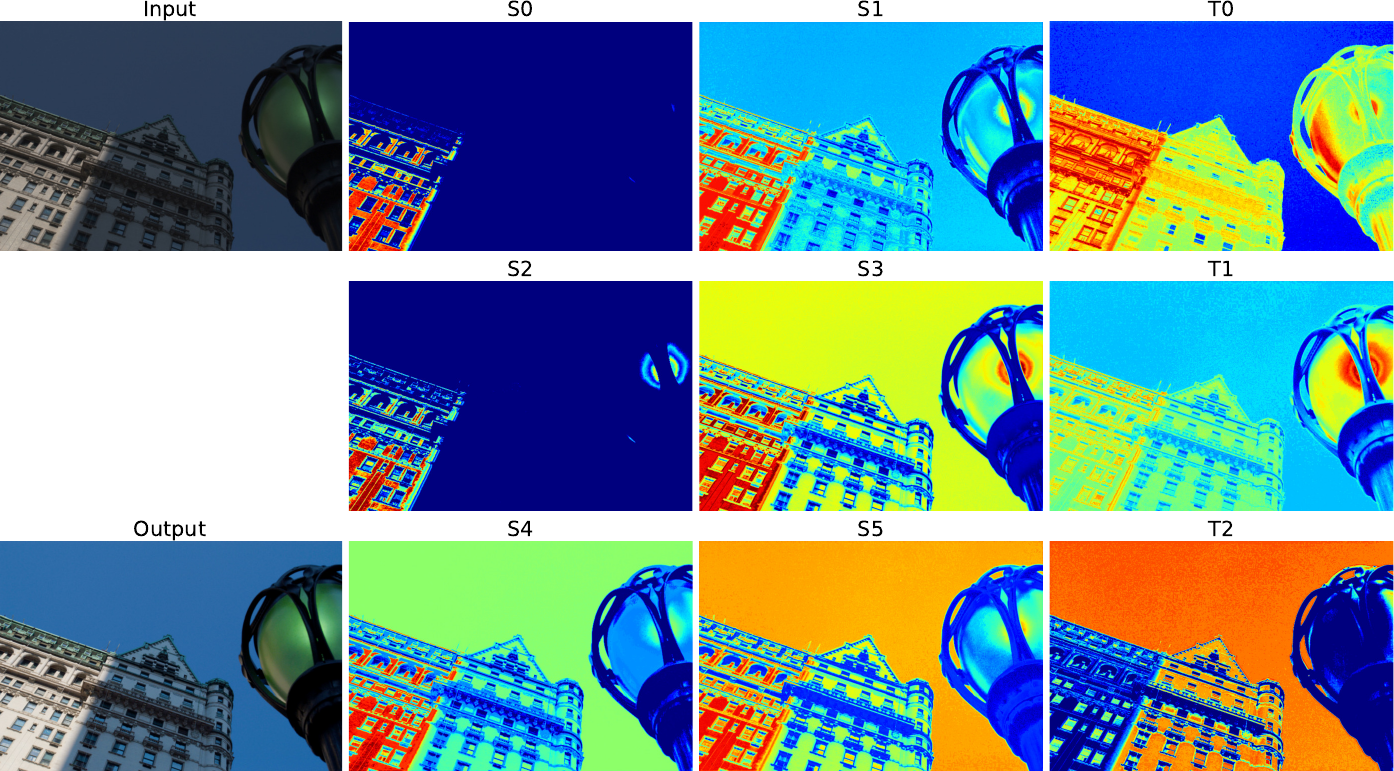}
        \caption{}
        \label{fig:analysis2}
    \end{subfigure}
  
    \caption{Visualization for results of the slicing and LUT transform. 
    The images in the first column are input and output image, respectively.
    The images in the second and third columns are results of slicing operation.
    The images in the forth column are results of LUT transform.
    The pixels closer to red indicate more activated, and the pixels closer to blue indicate less activated.  
    }  
    \label{fig:analysis}
\end{figure*}

\subsection{Loss Function}
\label{sec:loss}
Our loss function $\mathcal{L}$ comprises three terms as
\begin{equation}
    \mathcal{L} = \mathcal{L}_{mse} + \lambda_{c} \cdot \mathcal{L}_c + \lambda_{p} \cdot \mathcal{L}_p,
    \label{eq:loss}
\end{equation} 
which are the mean squared error (MSE), color difference loss, and perceptual loss, respectively. 
Specifically, $\mathcal{L}_{mse}$ calculates the MSE between predicted image $Y$ and ground truth, which ensures the fidelity of the enhanced image.
$\mathcal{L}_c$ is the distance on the CIE94 LAB space \cite{wang2021real}, and $\mathcal{L}_p$ is the LPIPS loss \cite{zhang2018unreasonable} on a pre-trained AlexNet \cite{krizhevsky2012imagenet} for the perceptual quality.
Also, $\lambda_c$ and $\lambda_p$ are set to 0.005 and 0.05 following the previous work \cite{kim2024image,wang2021real}, respectively.

\section{Experiments}
\subsection{Datasets}
We conduct experiments on FiveK \cite{bychkovsky2011learning} and PPR10K \cite{liang2021ppr10k} datasets.
The FiveK dataset contains 5,000 input images and five ground truths by five experts (A/B/C/D/E).
The FiveK dataset consists of two types of input image formats for each task: 8-bit sRGB for photo retouch and 16-bit XYZ for tone mapping task.
The ground truth has an 8-bit sRGB format that was retouched by expert C and is selected for comparison with previous LUT methods \cite{kim2024image,wang2021real,yang2022adaint,yang2022seplut,zeng2020lut}.
We split the dataset into 4,500 pairs for the training set and 500 pairs for the testing set, as provided split in \cite{zeng2020lut}.
We also use the 480p version \cite{zeng2020lut} to speed up in training stage.
The evaluation is carried out on 480p and 4K versions in the testing stage.
The PPR10K dataset contains 11,161 RAW portrait photos with three ground truths by three experts (a/b/c).
Input images and ground truths have 16-bit and 8-bit sRGB formats, respectively, for photo retouch tasks.
The PPR10K dataset is divided into 8,875 pairs for the training stage and 2,286 pairs for the test stage with the official split \cite{liang2021ppr10k}.
We separately conduct experiments for each ground truth on the 360p version.

\subsection{Application Settings}
We train our network in an end-to-end manner with standard Adam optimizer \cite{kingma2014adam} for 400 epochs.
The learning rate is initially $1 \times 10^{-4}$ and decayed by 0.1 factor every 100 epochs.
The minibatch size is 1 for FiveK and 16 for PPR10K.
In the experiments on the PPR10K dataset \cite{liang2021ppr10k}, we use the ResNet-18 \cite{he2016deep} as backbone $B(\cdot)$ for \cref{eq:backbone} and increase the rank by a factor of 2 on LUT and weight generator for fair comparison with previous work \cite{liang2021ppr10k,kim2024image,yang2022adaint,yang2022seplut}.
We implement our method on the PyTorch framework.
The slicing and LUT transform operation are implemented using CUDA extension to speed up.
The experiments are conducted on an NVIDIA Tesla V100 GPU except \cref{sec:efficiency}.

\begin{table*}[t]
  \begin{minipage}{\linewidth}
    \centering
    \caption{Quantitative comparisons of photo retouch on the FiveK dataset \cite{bychkovsky2011learning}.
    ``-'' means the result is not available due to insufficient GPU memory. 
    ``*'' indicates that the results are obtained from their paper, and ``/'' are absent results due to publicly unavailable code.
    The best and second-best results are in {\color{red}red} and {\color{blue}blue}, respectively.}
    \label{tab:quanta_fivek_pr} 
    \begin{tabular}{c|c|cccc|cccc}
    \hline
    \multirow{2}{*}{Method}         & \multirow{2}{*}{\#param} & \multicolumn{4}{c}{480p}        & \multicolumn{4}{c}{Full   Resolution (4K)} \\ \cline{3-10}
                                    &                          & PSNR  & SSIM  & $\Delta E_{ab}$ & Runtime ($ms$) & PSNR     & SSIM     & $\Delta E_{ab}$    & Runtime ($ms$)   \\ \hline
    UPE \cite{wang2019underexposed} & 927.1K                   & 21.88 & 0.853 & 10.80           & 4.27    & 21.65    & 0.859    & 11.09              & 56.88     \\
    DPE \cite{chen2018deep}         & 3.4M                     & 23.75 & 0.908 & 9.34            & 7.21    & -        & -        & -                  & -         \\
    HDRNet \cite{gharbi2017deep}    & 483.1K                   & 24.66 & 0.915 & 8.06            & 3.49    & 24.52    & 0.921    & 8.20               & 56.07     \\
    DeepLPF \cite{moran2020deeplpf} & 1.7M                     & 24.73 & 0.916 & 7.99            & 32.12   & -        & -        & -                  & -         \\
    CSRNet \cite{he2020conditional} & 36.4K                    & 25.19 & 0.925 & 7.76            & 3.09    & 24.82    & 0.924    & 7.94               & 77.10     \\
    3D LUT \cite{zeng2020lut}       & 593.5K                   & 25.29 & 0.923 & 7.55            & 1.02    & 25.25    & 0.932    & 7.59               & 1.04      \\
    SA-3DLUT* \cite{wang2021real}   & 4.5M                     & 25.50 & /     & /               & 2.27    & /        & /        & /                  & 4.39      \\
    SepLUT* \cite{yang2022seplut}   & 119.8K                   & 25.47 & 0.921 & 7.54            & 1.10    & 25.43    & 0.932    & 7.56               & 1.20      \\
    AdaInt  \cite{yang2022adaint}   & 619.7K                   & 25.49 & 0.926 &7.47             & 1.29    & 25.48    & 0.934    & {\color{blue}7.45}               & 1.59      \\
    SABLUT \cite{kim2024image}      & 463.7K                   & {\color{blue}25.66} & {\color{blue}0.930} & {\color{blue}7.29}            & 1.20    & {\color{blue}25.66}    & {\color{blue}0.937}    & {\color{red}7.27}               & 3.64      \\ \hline
    Ours                            & 160.5K                   & {\color{red}25.76} & {\color{red}0.931} & {\color{red}7.26}            & 1.37    & {\color{red}25.69}    & {\color{red}0.938}    & {\color{red}7.27}               & 1.38      \\ \hline
    \end{tabular}%
  \end{minipage}
  \begin{minipage}{.37\linewidth}
    \caption{Quantitative comparisons on the FiveK dataset \cite{bychkovsky2011learning} for tone mapping.}
    \centering
    \label{tab:quanta_fivek_tm}
    \begin{tabular}{c|ccc}
      \hline
      \multirow{2}{*}{Method}         & \multicolumn{3}{c}{480p} \\ \cline{2-4}
                                      & PSNR                 & SSIM                 &  $\Delta E_{ab}$     \\ \hline
      UPE \cite{wang2019underexposed} & 21.56                & 0.837                & 12.29  \\
      DPE \cite{chen2018deep}         & 22.93                & 0.894                & 11.09  \\
      HDRNet \cite{gharbi2017deep}    & 24.52                & 0.915                & 8.14   \\
      CSRNet \cite{he2020conditional} & 25.19                & 0.921                & 7.63   \\
      3DLUT \cite{zeng2020lut}        & 25.07                & 0.920                & 7.55   \\
      SepLUT* \cite{yang2022seplut}   & 25.43                & 0.922                & 7.43   \\
      AdaInt \cite{yang2022adaint}    & 25.28                & {\color{blue}0.925}        & 7.48  \\ 
      SABLUT \cite{kim2024image}      & {\color{blue}25.59}   & {\color{red}0.932}   & {\color{blue}7.14}   \\  \hline
      Ours                            & {\color{red}25.69}   & {\color{red}0.932}   & {\color{red}7.11}  \\\hline
    \end{tabular}%
  \end{minipage}%
  \hfill
  \begin{minipage}{.55\linewidth}
    \caption{Quantitative comparisons on the PPR10K dataset \cite{liang2021ppr10k} for photo retouch.}
    \centering
    \label{tab:quanta_ppr10k}
    \resizebox{\columnwidth}{!}{%
      \begin{tabular}{c|c|cccc}
      \hline
      Dataset                   & Method                          & PSNR                & $\Delta   E_{ab}$  & $PSNR^{HC}$          &  $\Delta E_{ab}^{HC}$  \\ \hline
      \multirow{5}{*}{PPR10K-a} & 3DLUT   \cite{liang2021ppr10k}  & 25.64               & 6.97               & 28.89                & 4.53                   \\
                                & SepLUT*   \cite{yang2022seplut} & 26.28               & 6.59               & /                    & /                      \\
                                & AdaInt   \cite{yang2022adaint}  & 26.33               & 6.56               & 29.57                & 4.26                   \\
                                & SABLUT \cite{kim2024image}      & {\color{blue}26.45}  & {\color{blue}6.51}& {\color{blue}29.70}  & {\color{blue}4.23}     \\
                                & Ours                            & {\color{red}26.50}  & {\color{red}6.46}  & {\color{red}29.74}   & {\color{red}4.20}      \\ \hline
      \multirow{5}{*}{PPR10K-b} & 3DLUT   \cite{liang2021ppr10k}  & 24.70               & 7.71               & 27.99                & 4.99                   \\
                                & SepLUT*   \cite{yang2022seplut} & 25.23               & 7.49               & /                    & /                      \\
                                & AdaInt   \cite{yang2022adaint}  & 25.40               & 7.33               & 28.65                & 4.75                   \\
                                & SABLUT \cite{kim2024image}      & {\color{blue}25.48} & {\color{red}7.19} & {\color{blue}28.72}  & {\color{red}4.66}       \\
                                & Ours                            & {\color{red}25.51}  & {\color{blue}7.23}  & {\color{red}28.75}   & {\color{blue}4.68}    \\ \hline
      \multirow{5}{*}{PPR10K-c} & 3DLUT   \cite{liang2021ppr10k}  & 25.18               & 7.58               & 28.49                & 4.92                   \\
                                & SepLUT*   \cite{yang2022seplut} & 25.59               & 7.51               & /                    & /                      \\ 
                                & AdaInt   \cite{yang2022adaint}  & 25.68               & 7.31               & 28.93                & 4.76                   \\
                                & SABLUT \cite{kim2024image}      & {\color{blue}25.72} & {\color{blue}7.28} & {\color{blue}29.00}  & {\color{blue}4.72}     \\ 
                                & Ours                            & {\color{red}25.77}  & {\color{red}7.25}  & {\color{red}29.04}   & {\color{red}4.71}      \\ \hline
      \end{tabular}%
    }
  \end{minipage} 
\end{table*}

\begin{table}[t]
\caption{Component-wise ablation study. ``+'' denotes adding LUTs or bilateral grids to match the model size of the original.}
\label{tab:comp_ablation}
\resizebox{\columnwidth}{!}{%
\begin{tabular}{c|ccccc}
\hline
              & PSNR  & SSIM  & $\Delta   E_{ab}$ & \#param  & FLOPs(M)     \\ \hline
Only Grid   & 25.21 & 0.926 & 7.62              & 112.872K & 81.38      \\
Only Grid+  & 25.41 & 0.927 & 7.50              & 159.924K & 154.96      \\
Only LUT    & 25.49 & 0.929 & 7.44              & 109.332K & 52.27       \\
Only LUT+   & 25.53 & 0.930 & 7.41              & 196.380K & 90.64      \\
1$\times$1 conv & 25.68 & 0.929 & 7.32       & 160.508K & 119.71       \\ \hline \hline
Ours        & 25.76 & 0.931 & 7.26              & 160.478K & 110.38        \\ \hline
\end{tabular}%
}
\end{table}

\subsection{The Component-wise Ablation Study}
The \cref{tab:comp_ablation} shows a component-wise ablation study on FiveK \cite{bychkovsky2011learning}.
Only Grid refers to evaluating results without the LUTs, and Only LUT does the same without the bilateral grids.
``+'' denotes adding LUTs or bilateral grids to match the number of parameters of the original.
The results indicate using LUTs and bilateral grids separately leads to suboptimal performance.
The ``1$\times$1 conv'' denotes using 1$\times$1 convolution to combine the spatial features, described in \cref{eq:fusion}.
In conjunction with \cref{tab:runtime}, 1$\times$1 convolution does not improve performance and inference time.

\subsection{Analysis of Slicing and LUT Transform}
In this section, we analyze the results of slicing and LUT transform.
For the analysis, we visualize the results as can be seen in \cref{fig:analysis}.
The results of slicing on the third column (S1, S3, S5) appear to globally adjust the input images.
In the results on the second column (S0, S2, S4), specific parts of the image are highlighted that seem to help with local adjustment.
Finally, the results of the LUT transform on the fourth column (T0, T1, T2) seem to adjust the correlation of colors.
Each result enhances the specific part and also can be further decomposed into distinct results by three 2D LUTs or 2D bilateral grids.
The weighted sum of these results provides the spatial-aware enhancement.
We demonstrate that our method can handle pixel-wise spatial-aware local enhancement by this analysis.

\begin{figure*}[tb]
  \centering
  \begin{subfigure}{\linewidth}
    \includegraphics[width=\columnwidth]{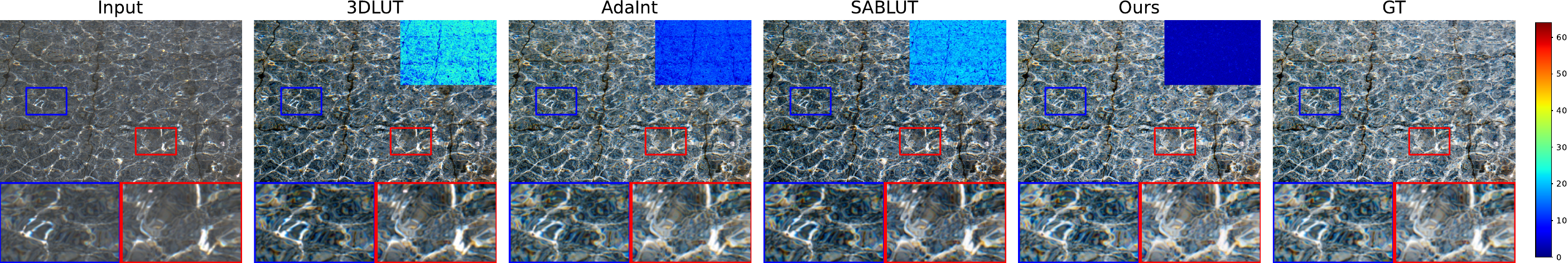}

    \label{fig:quali1}
  \end{subfigure}
  
  \begin{subfigure}{\linewidth}
    \includegraphics[width=\columnwidth]{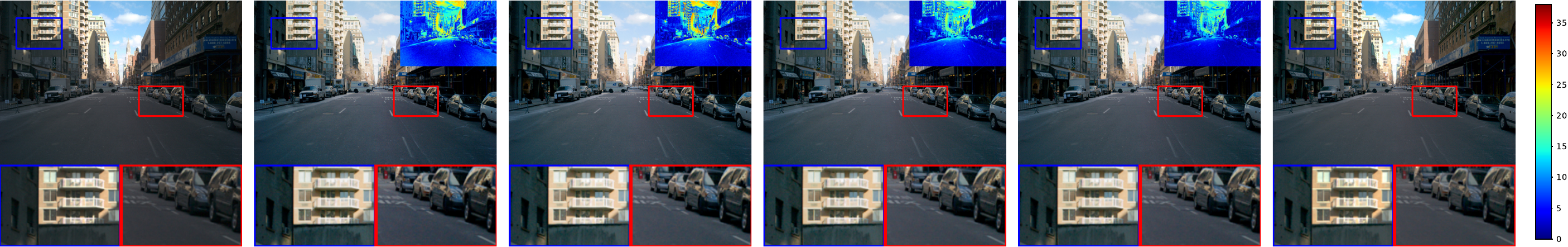}

    \label{fig:quali2}
  \end{subfigure}
  
  \begin{subfigure}{\linewidth}
    \includegraphics[width=\columnwidth]{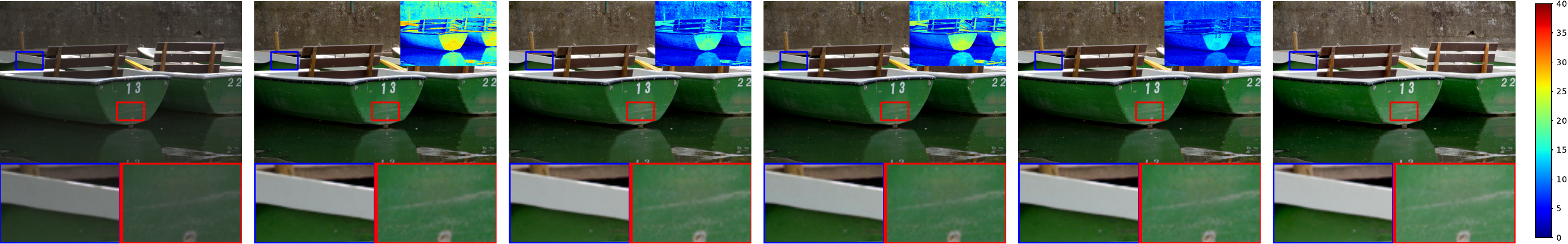}
    
    \label{fig:quali3}
  \end{subfigure} 
  
  \caption{Qualitative comparisons for photo retouch on the FiveK dataset \cite{bychkovsky2011learning}.
  The corresponding error maps on the upper right of each picture indicate differences with ground truth,
  where each color indicates the degree of error based on the corresponding color bars.
  } 
  \label{fig:quali}
\end{figure*}

\begin{table}[!tb]
\caption{Comparisons with efficient methods with CPU (i9-12900F) and cost-effective GPU$^{\dagger}$ (GTX 1660 SUPER) on the FiveK dataset \cite{bychkovsky2011learning} for photo retouch task.}

\label{tab:comp_with_effi}
\resizebox{\columnwidth}{!}{%
\begin{tabular}{c|c|cccc|c|cc}
\hline
\multirow{3}{*}{Method}     & \multirow{3}{*}{\#param} & \multirow{2}{*}{PSNR} & \multirow{2}{*}{SSIM} & \multirow{2}{*}{$\Delta   E_{ab}$} & \multirow{2}{*}{FLOPs}   & \multicolumn{3}{c}{Runtime ($ms$)}  \\  \cline{7-9}
                            &                          &                       &                       &                                    &                           & CPU                       & \multicolumn{2}{c}{GPU$^{\dagger}$} \\ \cline{3-9}
                            &                          & \multicolumn{4}{c|}{480p}                                                                                      & 480p                      & 480p                    & 4K        \\ \hline
NILUT \cite{conde2024nilut} & {\color{red}33.9K}       & 21.80                 & 0.879                 & 10.87                              & 11.59G                    & 182.84                    & 14.59                   & -         \\
ICELUT \cite{yang2024taming}& 299.3K                     & 25.27                 & 0.918                 & 7.51                               & {\color{red}15.02M}       & {\color{red}2.93}         & {\color{red}0.76}       & 2.75      \\
3DLUT \cite{zeng2020lut}    & 593.5K                   & 25.29                 & 0.923                 & 7.55                               & 92.52M                    & 20.87                     & 1.19                    & {\color{red}1.88}      \\
SABLUT \cite{kim2024image}  & 463.7K                   & 25.66                 & 0.930                 & 7.29                               & 70.11M                    & 26.8                      & 1.45                    & 11.97     \\ \hline \hline
Ours                        & 160.4K                   & {\color{red}25.76}    & {\color{red}0.931}    & {\color{red}7.26}                  & 110.37M                   & 21.71                     & 1.64                   & 3.94      \\ \hline
\end{tabular}%
}

\end{table}

\subsection{Comparison with State-of-the-Arts}
In this section, we compare our method with other SOTA real-time image enhancement methods \cite{wang2019underexposed,chen2018deep,gharbi2017deep,he2020conditional} and 3D LUT methods \cite{zeng2020lut,liang2021ppr10k,yang2022adaint,yang2022seplut,kim2024image}.
We conduct qualitative and quantitative comparisons on photo retouching and tone mapping tasks.
In the case of SA-3DLUT \cite{wang2021real} and SepLUT \cite{yang2022seplut} that do not provide publicly available code, the quantitative results are referenced from their paper, and the qualitative results are not provided.
 
\subsubsection{Quantitative Comparisons}
We compare the methods with PSNR, SSIM \cite{wang2004image}, and $L_2$-distance in CIE LAB color space ($\Delta E_{ab}$) for the FiveK dataset.
In comparisons on PPR10K, we additionally measure the human-centered PSNR ($PSNR^{HC}$) and color difference ($\Delta E^{HC}_{ab}$) \cite{liang2021ppr10k}.
For runtime comparison, we measure runtime on the same image set, by running 100 times and averaging them.
\cref{tab:quanta_fivek_pr} shows that our method outperforms other methods on both resolutions. 
Similar results can be seen in \cref{tab:quanta_fivek_tm} for the tone mapping task on the FiveK and \cref{tab:quanta_ppr10k} for the photo retouch task on the PPR10K.

Especially, our method overcomes the intrinsic problems of previous spatial-aware LUT methods, namely a tremendous number of parameters and a long inference time.
Our method has a much smaller size than SA-3DLUT \cite{wang2021real} (about 1/28 times) and SABLUT \cite{kim2024image} (about 1/3 times), which is comparable with non-spatial-aware 3D LUT methods \cite{yang2022adaint,yang2022seplut,zeng2020lut}.
Furthermore, our method provides fast inference not only at 480p but also at 4K resolution.

\subsubsection{Qualitative Comparisons}
\cref{fig:quali} provides several qualitative comparison results with corresponding error maps on the upper right of each picture.
The color of the error map represents the difference with ground truth based on the corresponding color bars on the right of each row.
The comparisons in the first row show that our method properly handles the color and texture.
Other methods produce an entirely darker color than the ground truth on both rectangles.
A noticeable difference can be seen in the tile colors.  
Specifically, our method produces a color which is much closer to ground truth.
The second row shows that spatial-aware methods (SABLUT, Ours) better handle the local enhancement than the others.
Especially, our method well produces the similar color with ground truth on both dark and bright areas.
The last row of the figure shows that our model has improved the overall color of the image. 
Specifically, the green color is closer to the ground truth than the other methods.

\subsection{Efficiency Analysis}
\label{sec:efficiency}
The \cref{tab:comp_with_effi} shows comparisons with efficient 3D LUT methods \cite{yang2024taming, conde2024nilut} on CPU (i9-12900F) and cost-effective GPU$^{\dagger}$ (GTX 1660 SUPER).
NILUT~\cite{conde2024nilut} adopts a neural implicit MLP in place of LUTs, and ICELUT~\cite{yang2024taming} enhances efficiency by replacing the backbone with LUTs.
Since NILUT failed to fit LUTs in 3DLUTs~\cite{zeng2020lut} for image-adaptiveness, we fitted NILUT to a randomly selected image.
For the others, we use settings in their official repositories.

Our method performs better than the other efficient methods and has reasonable inference time on both CPU and GPU$^{\dagger}$.
Although NILUT maintains a small number of parameters, its high FLOPs lead to longer inference time.
ICELUT \cite{yang2024taming} achieves efficient inference time, but its lack of spatial awareness results in suboptimal performance.
Although 2D LUT transform incurs a increase of FLOPs, our method shows only a minor difference compared to 3DLUT \cite{zeng2020lut} due to the shallow backbone. 
Furthermore, the cache-effective structure leads to a shorter inference time at 4K resolution than SABLUT \cite{wang2021real}.
When considered together with \cref{tab:quanta_fivek_pr}, we believe that the difference in inference time becomes wider as GPU performance decreases.

\section{Discussion and Conclusion}
We have developed an LUT-based image enhancement method, which decomposes the 3D LUT and 3D bilateral grid to reduce the model size and refine the spatial feature, providing a process that is cache-effective for short runtime.
We demonstrated that a 3D LUT can be decomposed by a combination of 2D LUTs and SVD, achieving nearly the same performance. 
We modified spatial feature fusion in a cache-efficient manner, which allows us to fast processing at high resolution. 
We have tested our method on two benchmark datasets, showing that it performs competitively compared to other state-of-the-art methods in terms of both performance and efficiency.

While our method guarantees real-time performance on GPUs, it might not achieve the same on low-end CPUs.
However, we believe that our method could be made applicable to low-end devices by adopting a shrink version as in NILUT or by replacing the backbone with LUTs, as done in ICELUT.
We also believe that the decomposition method and cache-efficient structure can be applied to other frameworks. These could be directions for future work.

\section*{Acknowledgements}
This work was partly supported by Institute of Information \& communications Technology Planning \& Evaluation (IITP) grant funded by the Korea government(MSIT) [NO.RS-2021-II211343, Artificial Intelligence Graduate School Program (Seoul National University)] and the BK21 FOUR program of the Education and Research Program for Future ICT Pioneers, Seoul National University in 2025.

{
    \small
    \bibliographystyle{ieeenat_fullname}
    \bibliography{main}
}
\appendix
\maketitlesupplementary
In this supplementary material, we provide the following.

\begin{itemize}
    \item \cref{sec:network_structure} Details of network structure.
    \item \cref{sec:analysis} Analysis for PPR10K Dataset.
    \item \cref{sec:bilinear} Details of 2D slicing and LUT transform.
    \item \cref{sec:additional_comp} Additional quantitative and qualitative comparisons.
\end{itemize}

\section{Details of Network Structure}
\label{sec:network_structure}
In this section, we provide the details of our network structure as described in \cref{tab:backbone,tab:gen_bilateral,tab:gen_bilateral_weight,tab:gen_lut,tab:gen_lut,tab:gen_lut_weight}.
The backbone in \cref{tab:backbone} consists of five layers of a convolutional neural network (CNN), with $m$ set to 8, following previous work \cite{kim2024image}.
Each generator in \cref{tab:gen_bilateral,tab:gen_bilateral_weight,tab:gen_lut,tab:gen_lut,tab:gen_lut_weight} is composed of two fully connected (FC) layers with the insight of rank factorization, which can reduce the parameters and training difficulty \cite{yang2022adaint,yang2022seplut,kim2024image}.
The bilateral grid generator $H_s(\cdot)$ in \cref{tab:gen_bilateral} generates 2D bilateral grids to replace 3D bilateral grids.
We decide the $D_s=17$, $K=6$ for fair comparison on spatial feature fusion with SABLUT \cite{kim2024image}.
The weights and biases for 2D bilateral grids are generated by the bilateral grid weight generator $H_{sw}(\cdot)$ in \cref{tab:gen_bilateral_weight}.
We experimentally select the number of hidden layers for the bilateral grid weight generator as $M_{sw} =8$.
Since three 2D bilateral grids replace a 3D bilateral grid, the number of weights $N_{sw}$ and biases $N_{sb}$ for each channel are 3 and 1, respectively.

The LUT generator $H_{t}(\cdot)$ in \cref{tab:gen_lut} generates the singular value decomposition (SVD) components of 2D LUTs with eight singular values $N_s=8$ as described in the main paper.
Most of previous 3D LUT methods \cite{liang2021ppr10k,kim2024image,wang2021real,yang2022adaint,yang2022seplut,zeng2020lut} have $D_t = 17$ or $D_t=33$ vertices.
When we set $D_t$ to 17, our model does not achieve the desired performance, with a PSNR of 25.54 dB on the photo retouch task on FiveK \cite{bychkovsky2011learning}.
Meanwhile, our method achieves a competitive performance of 25.76 dB with $D_t=33$ in the same task.
We decide to set $D_t$ as 33, based on the above experiment. 
The 2D LUTs can be easily reconstructed by matrix multiplication, as discussed in the main paper.
The LUT weight generator $H_{tw}(\cdot)$ in \cref{tab:gen_lut_weight} is similar to the bilateral grid weight generator. 
The parameters of the LUT weight generator, $M_{tw} = 8$, $N_{tw}=3$, and $N_{tb}=1$, are decided for the same reason as the bilateral grid weight generator.

\begin{table*}[tbh!]
    \centering
    \begin{minipage}{.35\linewidth}
      \caption{Details of the backbone network $B(\cdot)$ where $Conv3$ is $3 \times 3$ convolution block (stride=2, padding=1), $LR$ is LeakyReLU and $IN$ is instance normalization.}
      \label{tab:backbone}
      \centering
      \resizebox{\columnwidth}{!}{%
      \begin{tabular}{cc}
      \hline
      Layer                & Output Shape              \\ \hline
      Bilinear Downsample  & $3 \times 256 \times 256$ \\
      $Conv3 + LR + IN$    & $m \times 128 \times 128$ \\ 
      $Conv3 + LR + IN$    & $2m \times 64 \times 64$  \\ 
      $Conv3 + LR +IN$     & $4m \times 32 \times 32$  \\ 
      $Conv3 + LR +IN$     & $8m \times 16 \times 16$  \\ 
      $Conv3 + LR$         & $8m \times 8 \times 8$   \\ 
      Droupout (0.5)       & $8m \times 8 \times 8$   \\ 
      Average Pooling      & $8m \times 2 \times 2$   \\ 
      Reshape              & $32m$                     \\ \hline
      \end{tabular}
      }
    \end{minipage}
    \hfill
    \begin{minipage}{.64\linewidth}
      \centering
      \begin{minipage}{.54\linewidth}
        \centering
        \caption{Details of the bilateral grid generator $H_{s}(\cdot)$, where $FC$ is a fully connected layer. 
        $M_s$, $K$, and $D_{s}$ are the number of hidden layers, bilateral grids, and 
        bilateral grid vertices, respectively.}
        \label{tab:gen_bilateral}
        
        \begin{tabular}{cc}
        \hline
        Layer     & Output Shape                                           \\\hline
        $FC$      & $M_s$                                                  \\
        $FC$      & $K \cdot 3 \cdot D_{s}^2$                                    \\
        Reshape   & $K \times 3 \times D_{s} \times D_{s}$ \\\hline
        \end{tabular}
      \end{minipage}
      \hfill
      \begin{minipage}{.42\linewidth}
        \centering
        \caption{Details of the bilateral grid weight generator $H_{sw}(\cdot)$. 
        $M_{sw}$ is the number of hidden layers. $N_{sw}$ and $N_{sb}$ are the number of bilateral grid weights and biases for each channel, respectively.}
        \label{tab:gen_bilateral_weight}
        \resizebox{\columnwidth}{!}{%
        \begin{tabular}{cc}
        \hline
        Layer     & Output Shape                                           \\\hline
        $FC$      & $M_{sw}$                                                  \\
        $FC$      & $K \cdot 3 \cdot (N_{sw} + N_{sb}) $                                    \\
        Reshape   & $K \times 3 \times (N_{sw} + N_{sb})$ \\\hline
        \end{tabular}
        }
      \end{minipage}
  
      \begin{minipage}{.54\linewidth}
        \centering
        \caption{Details of the LUT generator $H_{t}(\cdot)$. $M_t$, $D_{t}$ and $N_{S}$ are the number of hidden layers, LUT vertices, and singular values, respectively.}
        \label{tab:gen_lut}
        \resizebox{\columnwidth}{!}{%
        \begin{tabular}{cc}
        \hline
        Layer     & Output Shape                                           \\\hline
        $FC$      & $M_t$                                                  \\
        $FC$      & $3 \cdot 3 \cdot (D_{t} \cdot N_{S} + N_{S} + D_{t} \cdot N_{S})$                                    \\
        Reshape   & \makecell{$U : 3 \times 3 \times (D_{t} \times N_{S})$ \\ 
                              $S : 3 \times 3 \times N_{S}$ \\
                              $V : 3 \times 3 \times (D_{t} \times N_{S})$}\\ \hline
        \end{tabular}
        }
      \end{minipage}
      \hfill
      \begin{minipage}{.42\linewidth}
        \centering
        \caption{Details of the LUT weight generator $H_{tw}(\cdot)$. 
        $M_{tw}$ are the number of hidden layers. $N_{tw}$ and $N_{tb}$ are the number of LUT weights and biases for each channel, respectively.}
        \label{tab:gen_lut_weight}
        \resizebox{\columnwidth}{!}{%
        \begin{tabular}{cc}
        \hline
        Layer     & Output Shape                                           \\\hline
        $FC$      & $M_{tw}$                                                  \\
        $FC$      & $3 \cdot 3 \cdot (N_{tw} + N_{tb}) $                                    \\
        Reshape   & $3 \times 3 \times (N_{tw} + N_{tb})$ \\\hline
        \end{tabular}
        }
      \end{minipage}
    \end{minipage}  
  \end{table*}
\begin{figure}[tb]
  \centering
  \begin{subfigure}{0.49\linewidth}
      \includegraphics[width=\columnwidth]{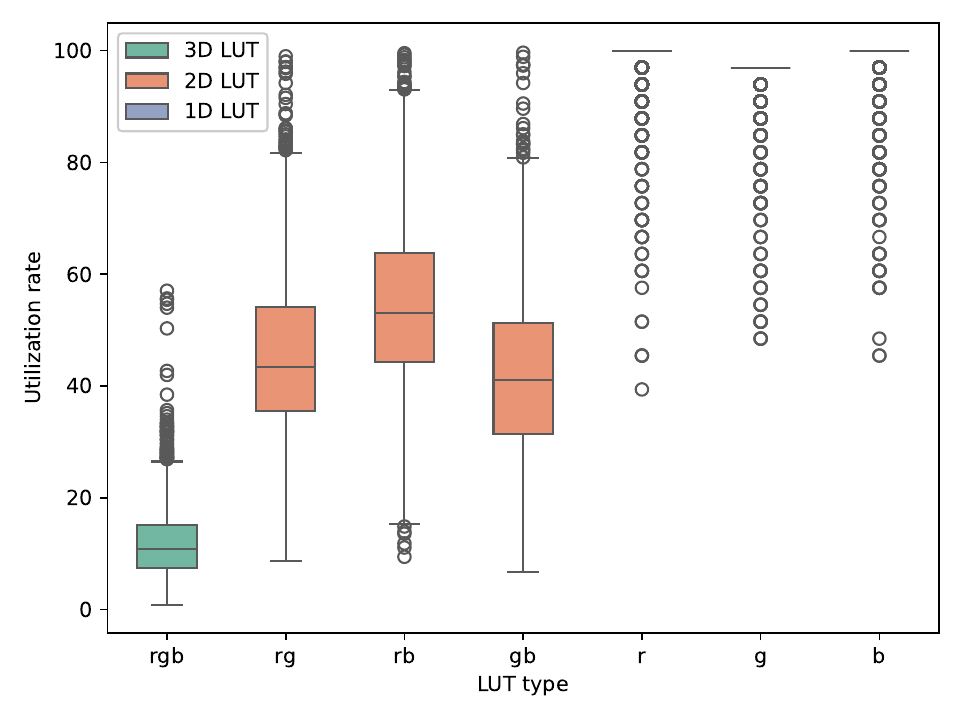}
      \caption{LUT utilization rate}
      \label{fig:lut_util_ppr10k}
  \end{subfigure}
  \hfill
  \begin{subfigure}{0.49\linewidth}
      \includegraphics[width=\columnwidth]{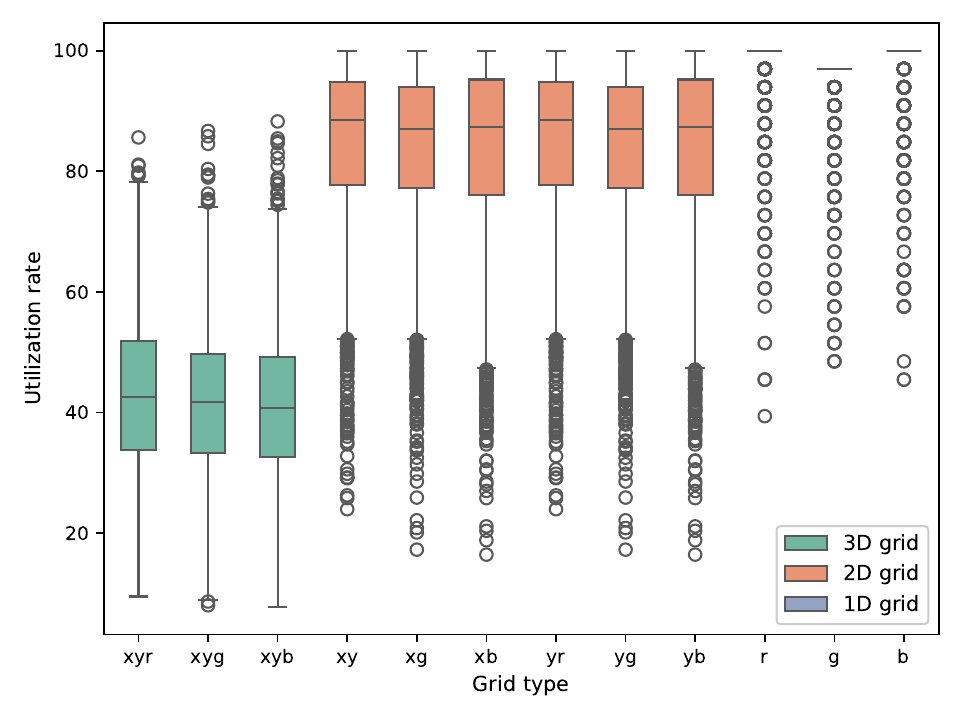}
      \caption{Grid utilization rate}
      \label{fig:grid_util_ppr10k}
  \end{subfigure}

  \begin{subfigure}{0.49\linewidth}
      \includegraphics[width=\columnwidth]{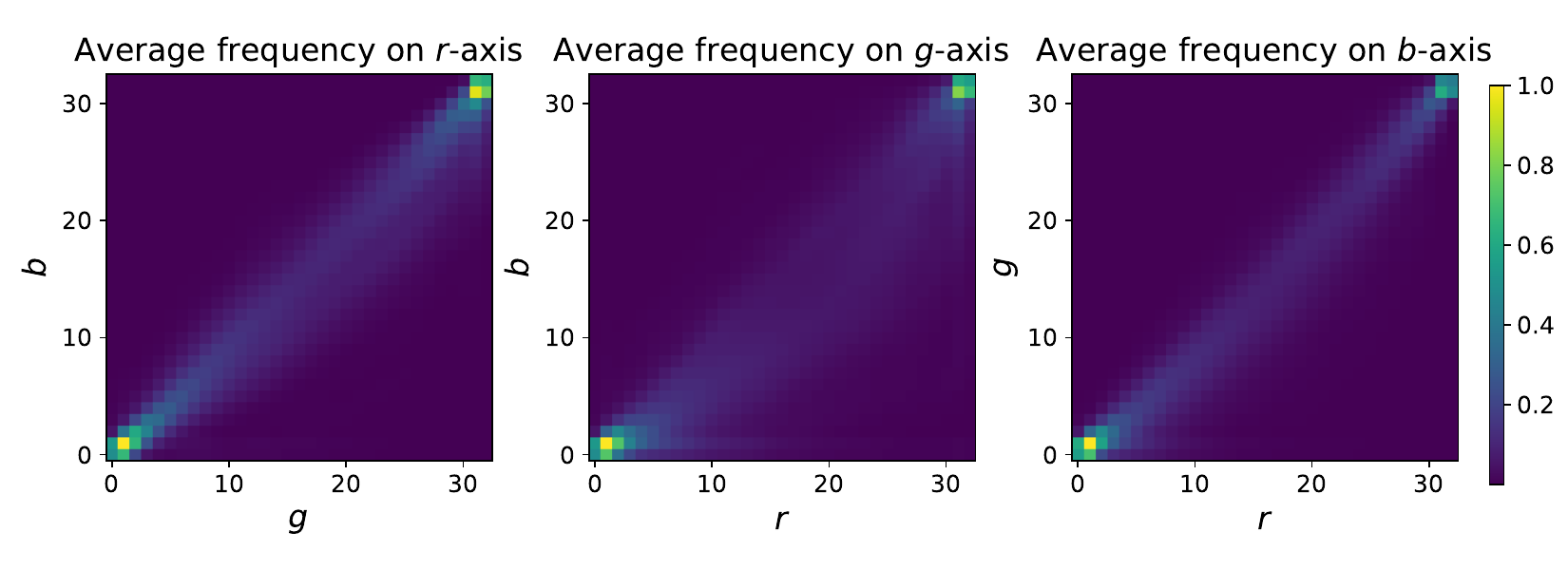}
      \caption{LUT occurrence statistics}
      \label{fig:lut_occurence_ppr10k}
  \end{subfigure}
  \hfill
  \begin{subfigure}{0.49\linewidth}
      \includegraphics[width=\columnwidth]{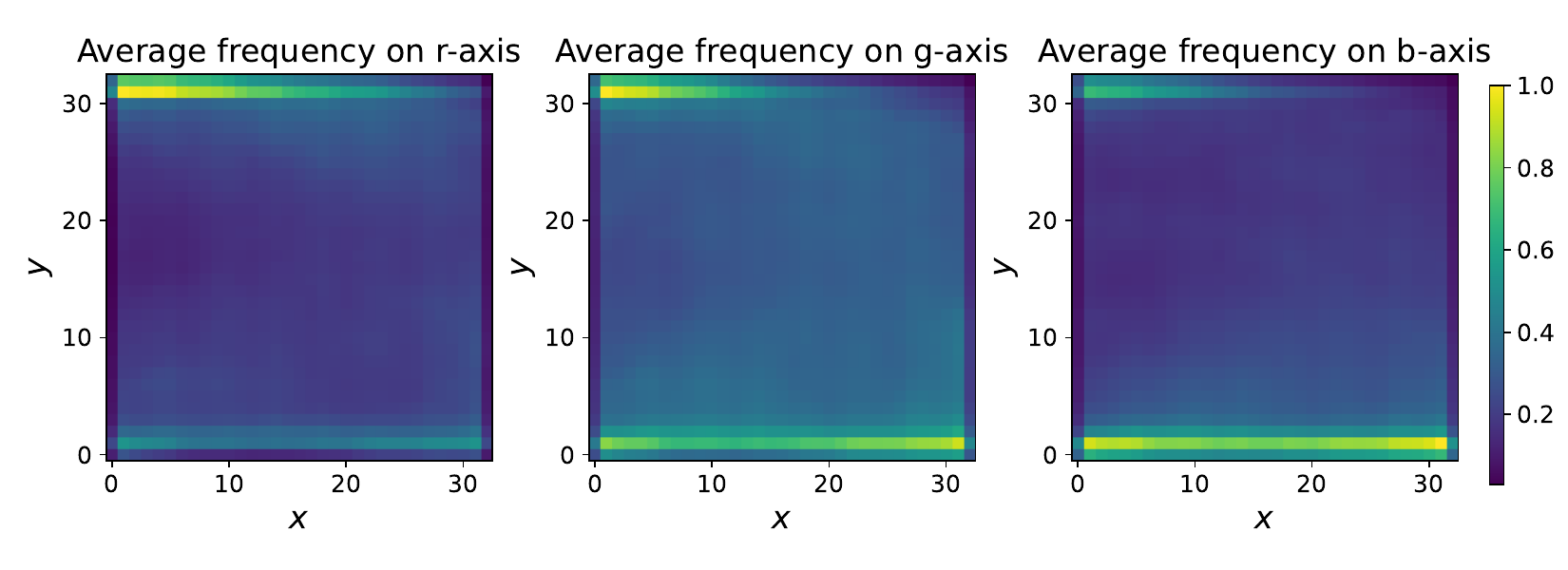}
      \caption{Grid occurrence statistics}
      \label{fig:grid_occurence_ppr10k}
  \end{subfigure}

  \caption{(a) and (b) are box plots of utilization rates on PPR10K dataset \cite{liang2021ppr10k} under different dimensions of LUT and bilateral grid. 
          Each color represents the dimension of LUT and bilateral grid. 
          (c) and (d) are LUT visualizations of average occurrence statistics on each axis for PPR10K.
          The cells closer to yellow indicate more frequently accessed vertices and the cells closer to blue indicate less frequently accessed vertices.}
  \label{fig:util_rate_ppr10k}  
\end{figure}

\section{Analysis for PPR10K Datasets}
\label{sec:analysis}
We also conduct an analysis for PPR10K on utilization rates and occurrence statistics.
The utilization rate indicates how many vertices of a 3D LUT are referenced compared to generated vertices for each image, like $\frac{\# referenced \ vertices}{\# generated \ vertices} \times 100$.
The occurrence statistics are estimated by counting the number of accesses for each vertex.
The overall tendency is similar to the results of the analysis for FiveK in the main paper. 
\cref{fig:lut_util_ppr10k} presents LUT utilization rate of PPR10K.
The rate of 3D LUT is very low, and 1D LUT is saturated.
As can be seen in \cref{fig:lut_occurence_ppr10k}, the occurrence statistics are concentrated on the diagonal.
Notably, the vertices near (1,1) and (32,32) are accessed more frequently than other diagonal vertices, compared to the FiveK dataset.
The bilateral grid has a broader distribution than LUT, but it also has a similar tendency.
The 3D bilateral grid is redundant, and the 1D bilateral grid is saturated, similar to the FiveK dataset, as shown in \cref{fig:grid_util_ppr10k}.
The vertices access is concentrated on as visualized in the first row and the thirty-second row in \cref{fig:grid_occurence_ppr10k}.
\begin{figure*}[tb]
    \centering
    
    \begin{subfigure}{\linewidth} 
      \includegraphics[width=\columnwidth]{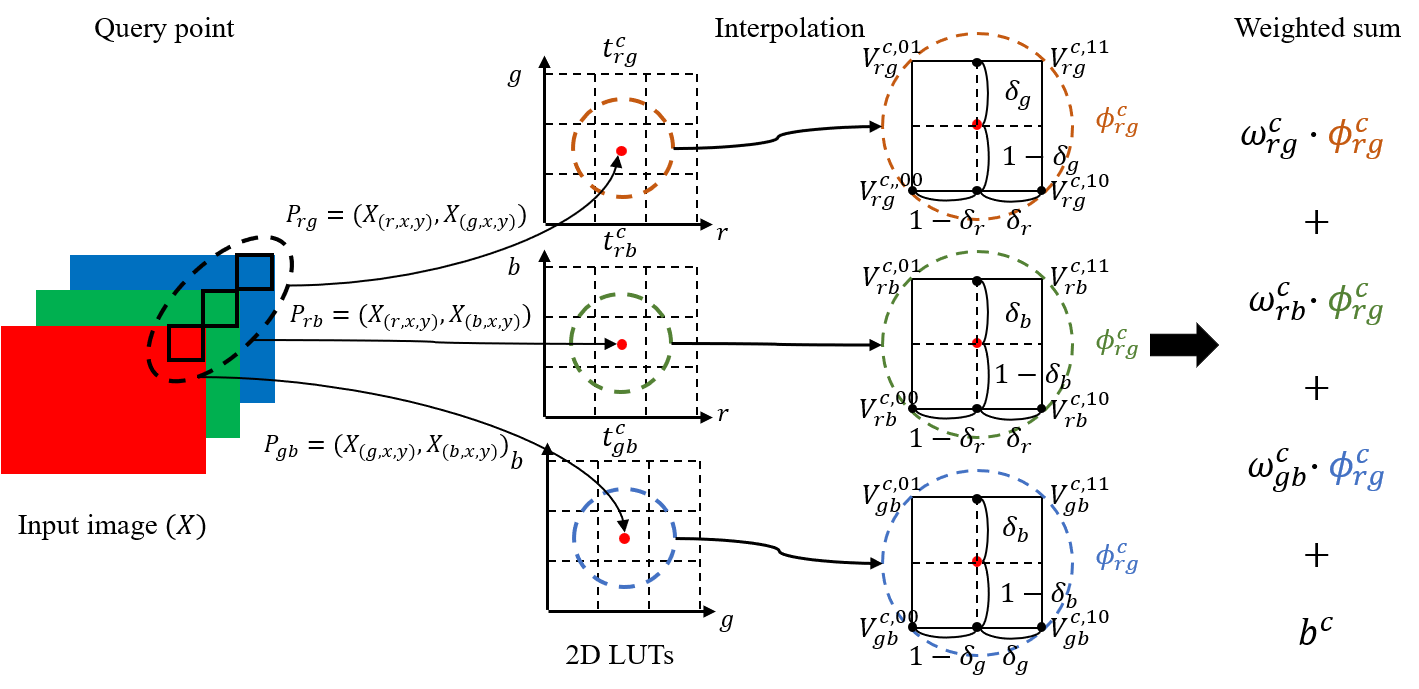}
      \caption{Description of the 2D LUT transform}
     
      \label{fig:bilinear_lut}
    \end{subfigure}
    \vfill
    \begin{subfigure}{\linewidth} 
      \includegraphics[width=\columnwidth]{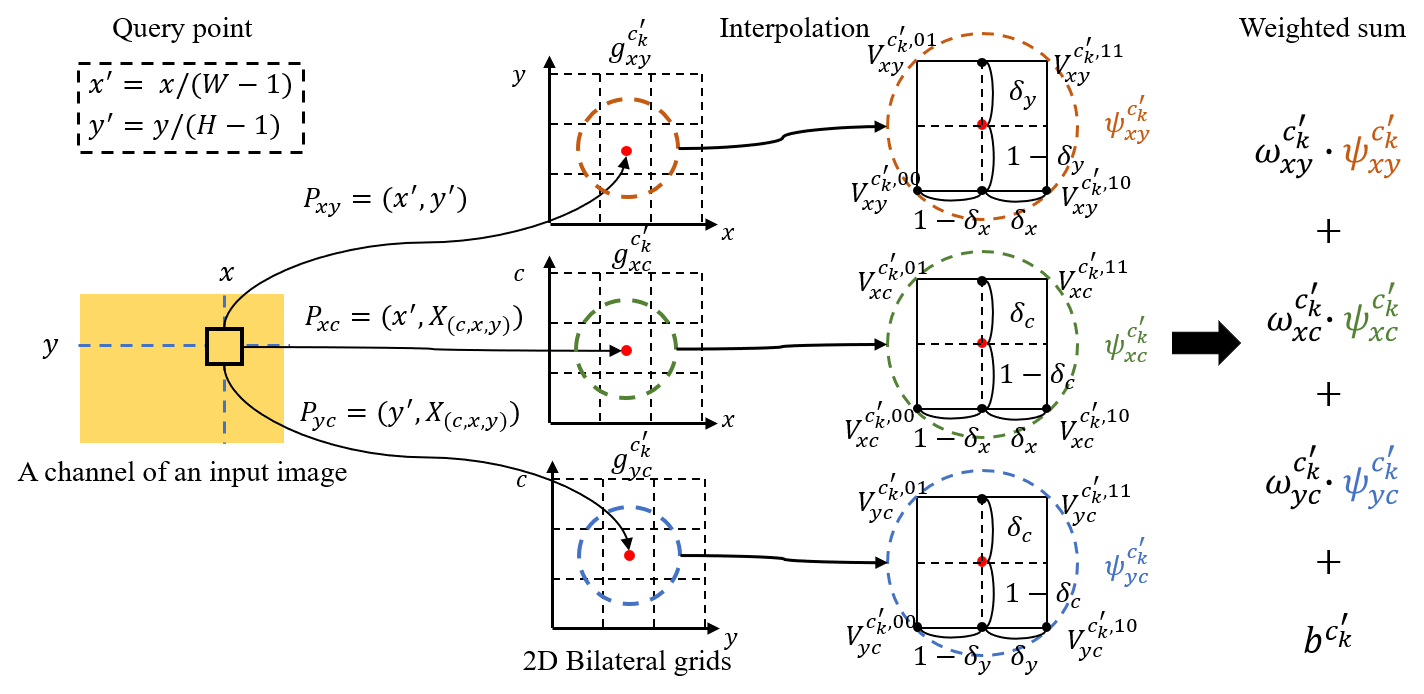}
      \caption{Description of the 2D slicing}
     
      \label{fig:bilinear_grid}
    \end{subfigure}
      
    \caption{Detailed description of 2D LUT transform (a) and 2D Slicing (b) based on bilinear interpolation.} 
    \label{fig:bilinear}
  \end{figure*}
  
  \section{Details of 2D Slicing and LUT Transform} 
  \label{sec:bilinear}
 In this section, we provide a detailed description of the 2D LUT Transform and 2D Slicing.
  \subsection{2D LUT Transform}
  \cref{fig:bilinear_lut} illustrates the detailed operation of 2D LUT transform $Transform_{2D}^{c}(X_{(c,x,y)},T^{2D})$ based on bilinear interpolation, where $T^{2D} \in \{t^{c}_{rg},t^{c}_{rb},t^{c}_{gb}|c \in \{r,g,b\}\}$ represents 2D LUTs and $X_{(c,x,y)}$ is the pixel value on $(c,x,y)$.
 The 2D LUT transform comprises the following three steps.
 First of all, query points for 2D LUTs are found based on input color values, which can be described as
  \begin{equation}
    \label{eq:query_lut}
    \begin{split}
      P_{rg}& = (p_r, p_g) = (X_{(r,x,y)}, X_{(g,x,y)}), \\
      P_{rb}& = (p_r, p_b) = (X_{(r,x,y)}, X_{(b,x,y)}), \\
      P_{gb}& = (p_g, p_b) = (X_{(g,x,y)}, X_{(b,x,y)}),     
    \end{split}
  \end{equation}
  where $P_{rg}$, $P_{rb}$, and $P_{gb}$ are query points for $t^c_{rg}$, $t^c_{rb}$, and $t^c_{gb}$, respectively. 
  
  Second, the bilinear interpolation is carried out to calculate retrieved values $\phi^c_{\alpha \beta}$ from $t^c_{\alpha \beta}$ with $\alpha \beta \in \{rg,rb,gb\}$.
  For the interpolation operation, we find the left and right vertices of each query point on their axis. The left and right vertices can be denoted as
  \begin{equation}
    \label{eq:lookup}
    \begin{split}
      p^l_{\alpha}& = \lfloor p_{\alpha} \cdot (D_{t}-1)\rfloor / (D_{t}-1), \\
      p^r_{\alpha}& = p^l_{\alpha} + 1 / (D_{t}-1), \\
      p^l_{\beta}& = \lfloor p_{\beta} \cdot (D_{t}-1)\rfloor / (D_{t}-1), \\
      p^r_{\beta}& = p^l_{\beta} + 1 / (D_{t}-1), 
    \end{split}
  \end{equation}
  where $\lfloor \cdot \rfloor$ is floor operator.
  The four adjacent points on a 2D LUT can be found as
  \begin{equation}
    \label{eq:adjacent}
    \begin{split}
      V^{c,00}_{\alpha \beta}& = t^{c}_{\alpha \beta}(p^l_{\alpha}, p^l_{\beta}), \\
      V^{c,10}_{\alpha \beta}& = t^{c}_{\alpha \beta}(p^r_{\alpha}, p^l_{\beta}), \\
      V^{c,01}_{\alpha \beta}& = t^{c}_{\alpha \beta}(p^l_{\alpha}, p^r_{\beta}), \\
      V^{c,11}_{\alpha \beta}& = t^{c}_{\alpha \beta}(p^r_{\alpha}, p^r_{\beta}).
    \end{split}
  \end{equation}
  The retrieved values can be calculated by bilinear interpolation, which can be formulated as
  \begin{equation}
    \label{eq:bilinear}
    \begin{split}
      \phi^c_{\alpha \beta, (x,y)}& = (1 - \delta_{\alpha}) \cdot (1 - \delta_{\beta}) \cdot V^{c,00}_{\alpha \beta} \\
                                 & + \delta_{\alpha} \cdot (1 - \delta_{\beta}) \cdot V^{c,10}_{\alpha \beta} \\
                                 & + (1 - \delta_{\alpha}) \cdot \delta_{\beta} \cdot V^{c,01}_{\alpha \beta} \\
                                 & + \delta_{\alpha} \cdot \delta_{\beta} \cdot V^{c,11}_{\alpha \beta}, 
    \end{split}
  \end{equation}
  where $\delta_{\alpha} = (p_{\alpha} - p^l_{\alpha})/(p^r_{\alpha} - p^l_{\alpha})$ and $\delta_{\beta} = (p_{\beta} - p^l_{\beta})/(p^r_{\beta} - p^l_{\beta})$.
  
  Finally, the weighted sum is conducted to calculate output values $\Phi^c$ of a 2D LUT transform like
  \begin{equation}
    \label{eq:weighte_sum}
    \begin{split}
      \Phi^c_{(x,y)}& = w^c_{rg} \cdot \phi^c_{rg,(x,y)}  \\
                    & + w^c_{rb} \cdot \phi^c_{rb, (x,y)}  \\
                    & + w^c_{rb} \cdot \phi^c_{rb, (x,y)} + b^c,
    \end{split}
  \end{equation}
  where $w^c_{rg}$, $w^c_{rb}$, $w^c_{gb}$, and $b^c$ are generated weights by LUT weight generator $H_{tw}(\cdot)$ in \cref{sec:network_structure}.
  
  \subsection{2D Slicing}
 As can be seen in \cref{fig:bilinear_grid}, the 2D slicing operation $Slicing_{2D}^{c'_k}(X, G^{2D})$ is similar to the LUT transform since both operations are based on bilinear interpolation.
 First, we find the query points based on the spatial coordinate $(x,y)$ and the corresponding color value on each channel of the image, which can be denoted as
  \begin{equation}
    \label{eq:query_grid}
    \begin{split}
      P_{xy}& = (p_x, p_y) = (x', y'), \\
      P_{xc}& = (p_x, p_c) = (x', X_{(c'_{k'},x,y)}), \\
      P_{yc}& = (p_y, p_c) = (y', X_{(c'_{k'},x,y)}),     
    \end{split}
  \end{equation}
 where $x'= x/(W-1)$, $y'= y/(H-1)$ and $k' = mod(k, 3)$.
 
 Second, the bilinear interpolation is also carried out on 2D bilateral grids $g^{c'_k}_{\alpha \beta}$ for retrieved values $\psi^{c'_k}_{\alpha \beta}$ with $\alpha \beta \in \{xy,xc,yc\}$.
 The neighboring vertices and adjacent points of each query point for slicing can be defined as
 \begin{equation}
  \label{eq:lookup_grid}
  \begin{split}
    p^l_{\alpha}& = \lfloor p_{\alpha} \cdot (D_{s}-1)\rfloor / (D_{s}-1), \\
    p^r_{\alpha}& =  p^l_{\alpha} + 1 / (D_{s}-1), \\
    p^l_{\beta}& = \lfloor p_{\beta} \cdot (D_{s}-1)\rfloor / (D_{s}-1), \\
    p^r_{\beta}& =  p^l_{\beta} + 1 / (D_{s}-1),
  \end{split}
\end{equation}
 
 \begin{equation}
  \label{eq:adjacent_grid}
  \begin{split}
    V^{c'_k,00}_{\alpha \beta}& = g^{c'_k}_{\alpha \beta}(p^l_{\alpha}, p^l_{\beta}), \\
    V^{c'_k,10}_{\alpha \beta}& = g^{c'_k}_{\alpha \beta}(p^r_{\alpha}, p^l_{\beta}), \\
    V^{c'_k,01}_{\alpha \beta}& = g^{c'_k}_{\alpha \beta}(p^l_{\alpha}, p^r_{\beta}), \\
    V^{c'_k,11}_{\alpha \beta}& = g^{c'_k}_{\alpha \beta}(p^r_{\alpha}, p^r_{\beta}).
  \end{split}
\end{equation}
The retrieved values by slicing can be formulated as
\begin{equation}
  \label{eq:bilinear_grid}
  \begin{split}
    \psi^{c'_k}_{\alpha \beta, (x,y)}& = (1 - \delta_{\alpha}) \cdot (1 - \delta_{\beta}) \cdot V^{c'_k,00}_{\alpha \beta} \\
                                     & + \delta_{\alpha} \cdot (1 - \delta_{\beta}) \cdot V^{c'_k,10}_{\alpha \beta} \\
                                     & + (1 - \delta_{\alpha}) \cdot \delta_{\beta} \cdot V^{c'_k,01}_{\alpha \beta} \\
                                     & + \delta_{\alpha} \cdot \delta_{\beta} \cdot V^{c'_k,11}_{\alpha \beta}. 
  \end{split}
  \end{equation}

 Finally, output values $\Psi^{c'_k}$ of 2D slicing can be calculated through a weighted sum with generated weights by the bilateral grid weight generator $H_{sw}(\cdot)$ in \cref{sec:network_structure}.
 The weighted sum can be denoted as
  \begin{equation}
    \label{eq:weighte_sum_gri1d}
    \begin{split}
      \Psi^{c'_k}_{(x,y)}& = w^{c'_k}_{xy} \cdot \psi^{c'_k}_{xy, (x,y)} \\ 
                         & + w^{c'_k}_{xc} \cdot \psi^{c'_k}_{xc, (x,y)} \\
                         & + w^{c'_k}_{yc} \cdot \psi^{c'_k}_{yc, (x,y)} + b^{c'_k}.
    \end{split}
  \end{equation}

 \subsection{Slicing and LUT Transform}
 Using notations in previous sections, the cache-effective slicing and LUT transform in the main paper can be rewritten as
 \begin{equation}
  \label{eq:slicing_transform}
  \begin{split}
    Y_{(r,x,y)}& = \Phi^r_{(x,y)} + \sum_{k=0}^{K/3-1}\Psi^{3k}_{(x,y)}, \\
    Y_{(g,x,y)}& = \Phi^g_{(x,y)} + \sum_{k=0}^{K/3-1}\Psi^{1 + 3k}_{(x,y)}, \\
    Y_{(b,x,y)}& = \Phi^b_{(x,y)} + \sum_{k=0}^{K/3-1}\Psi^{2 + 3k}_{(x,y)}.
  \end{split}
 \end{equation}

\section{Additional Quantitative and Qualitative Comparisons}
\label{sec:additional_comp}
In this section, we provide additional quantitative comparisons on HDRTV1K dataset \cite{chen2021new}.
Additional qualitative comparisons are conducted on FiveK \cite{bychkovsky2011learning}, PPR10K \cite{liang2021ppr10k}, and HDRTV1K \cite{chen2021new}.

\subsection{Additional Dataset}
The HDRTV1K is a dataset for the SDRTV-to-HDRTV task, which converts SDR contents to their HDRTV version. 
This dataset comprises captured images from 22 HDR10 videos and their corresponding SDR versions.
All HDR10 videos are encoded using PQ-OETF and the rec.2020 gamut.
The 1235 images from 18 videos are used in the training stage, and 117 images from 4 videos are used in the testing stage. 

\begin{table}[!htb]
\caption{Quantitative comparisons on HDRTV1K \cite{chen2021new}. The best and second-best results are in {\color{red}red} and {\color{blue}blue}, respectively.}
\label{tab:comp_video}
\resizebox{\columnwidth}{!}{%
\begin{tabular}{c|ccccc}
\hline
Method                          & PSNR               & SSIM                 & $\Delta E_{ITP}$    & HDR-VPD3              & Runtime(ms)   \\ \hline
HDRNet \cite{gharbi2017deep}    & 35.73              & {\color{red}0.9664}  & 11.52               & 8.462                 & 56.07       \\
CSRNet \cite{he2020conditional} & 35.04              & 0.9625               & 14.28               & 8.400                 & 77.1        \\
3DLUT  \cite{zeng2020lut}       & 36.06              & 0.9609               & 10.73               & 8.353                 & {\color{red}1.04}        \\
AdaInt \cite{yang2022adaint}    & 36.22              & 0.9658               & 10.89               & 8.423                 & 1.59        \\
SABLUT \cite{kim2024image}      & 36.41              & 0.9657               & 10.28               & 8.460                 & 3.64        \\
HDRTVNet \cite{chen2021new}       & {\color{red}36.88} & 0.9655               & {\color{red}9.78}   & {\color{blue}8.464}   & 70.01       \\ \hline \hline
Ours                            & {\color{blue}36.74}& {\color{blue}0.9663} & {\color{blue}9.99}  & {\color{red}8.500}    & {\color{blue}1.38}        \\ \hline
\end{tabular}%
}
\end{table}

\subsection{Additional Quantitative Comparisons}
We compare our method with other SOTA real-time methods \cite{gharbi2017deep, he2020conditional, zeng2020lut, yang2022adaint,kim2024image} and HDRTVNet \cite{chen2021new} on the HDRTV1K \cite{chen2021new}. 
HDRTVNet is a method for the SDRTV-to-HDRTV task, which is introduced together with the HDRTV1K dataset in their paper \cite{chen2021new}.
HDRTVNet is set to the fastest configuration to compare the real-time performance, which only uses the adaptive global color mapping.

We measure PSNR, SSIM, $\Delta E_{ITP}$ \cite{ITU-R_BT2124}, and HDR-VPD3 \cite{mantiuk2011hdr}.
The $\Delta E_{ITP}$ is the color difference on the ICtCp space and is designed for HDRTV.
HDR-VDP3 is an improved version of HDR-VDP2 that supports the rec.2020 gamut.
We measure these metrics using codes in the official repository of HDRNet.
As the HDRTV1K dataset was captured from video sequences, this experiment can offer insights into the performance on video.

As can be seen in \cref{tab:comp_video}, our method shows suitable performance and inference time on real-time video processing.
Although HDRTVNet has the best score on PSNR and $\Delta E_{ITP}$, it fails to achieve real-time performance under the fastest configuration.
Our model delivers real-time processing with a minor performance drop: 0.11 dB on PSNR, 0.0001 on SSIM, and 0.11 on $\Delta E_{ITP}$.

\subsection{Additional Qualitative Comparisons}
\label{sec:another_qual}
\begin{figure*}[tb]
    \centering
    \begin{subfigure}{0.85\linewidth} 
      \includegraphics[width=\columnwidth]{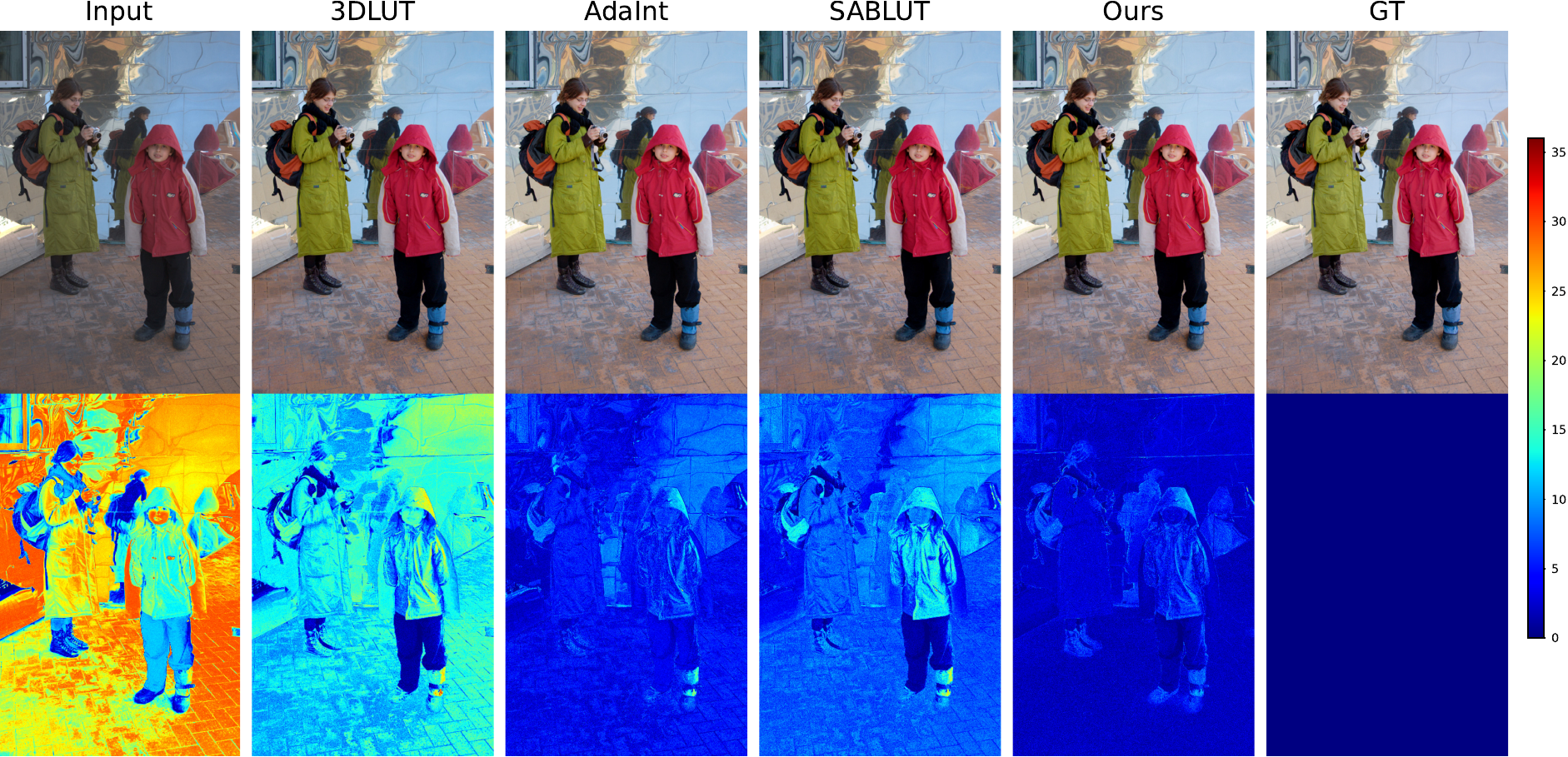}
      \label{fig:quali_fivek1}
    \end{subfigure}
    
    \begin{subfigure}{0.85\linewidth}
      \includegraphics[width=\columnwidth]{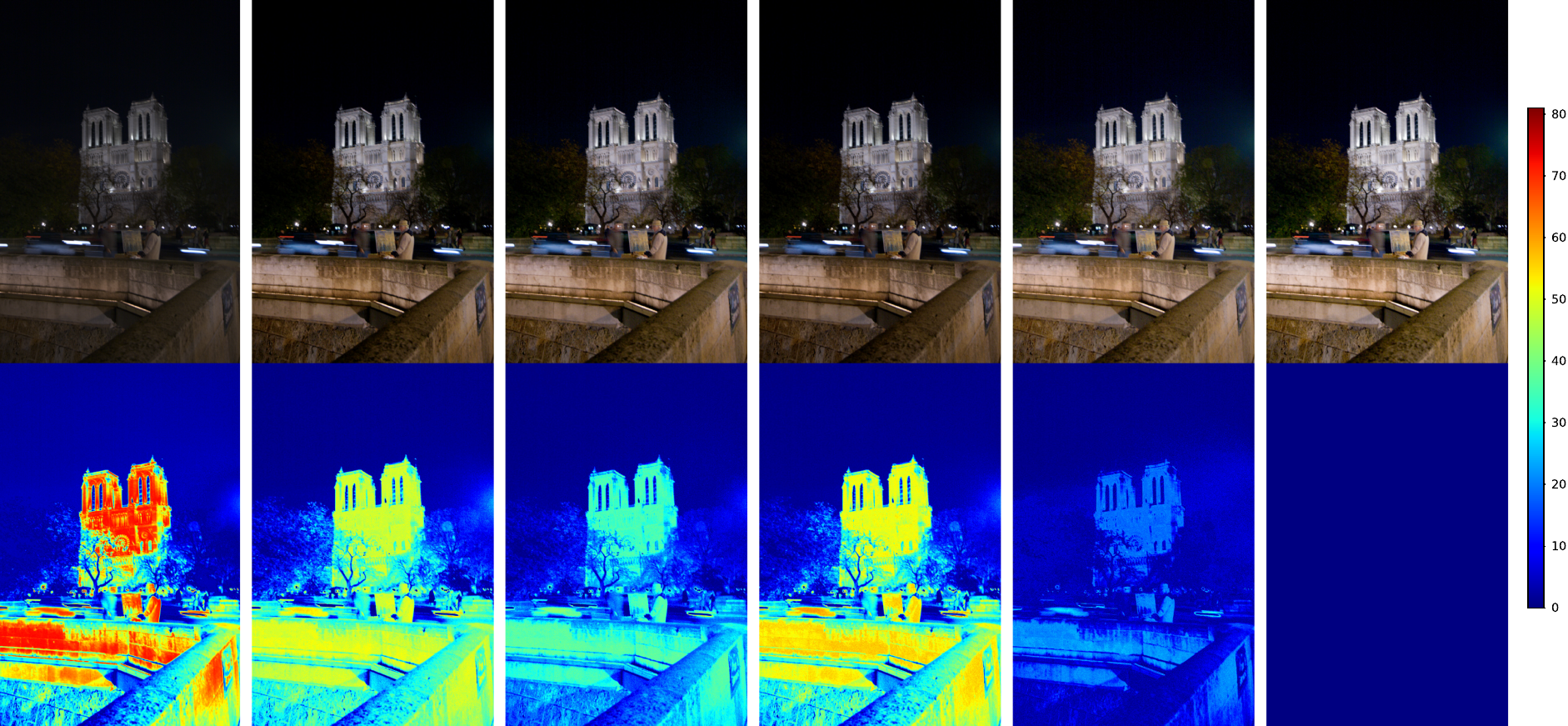}
      \label{fig:quali_fivek2}
    \end{subfigure}
    
    \begin{subfigure}{0.85\linewidth}
      \includegraphics[width=\columnwidth]{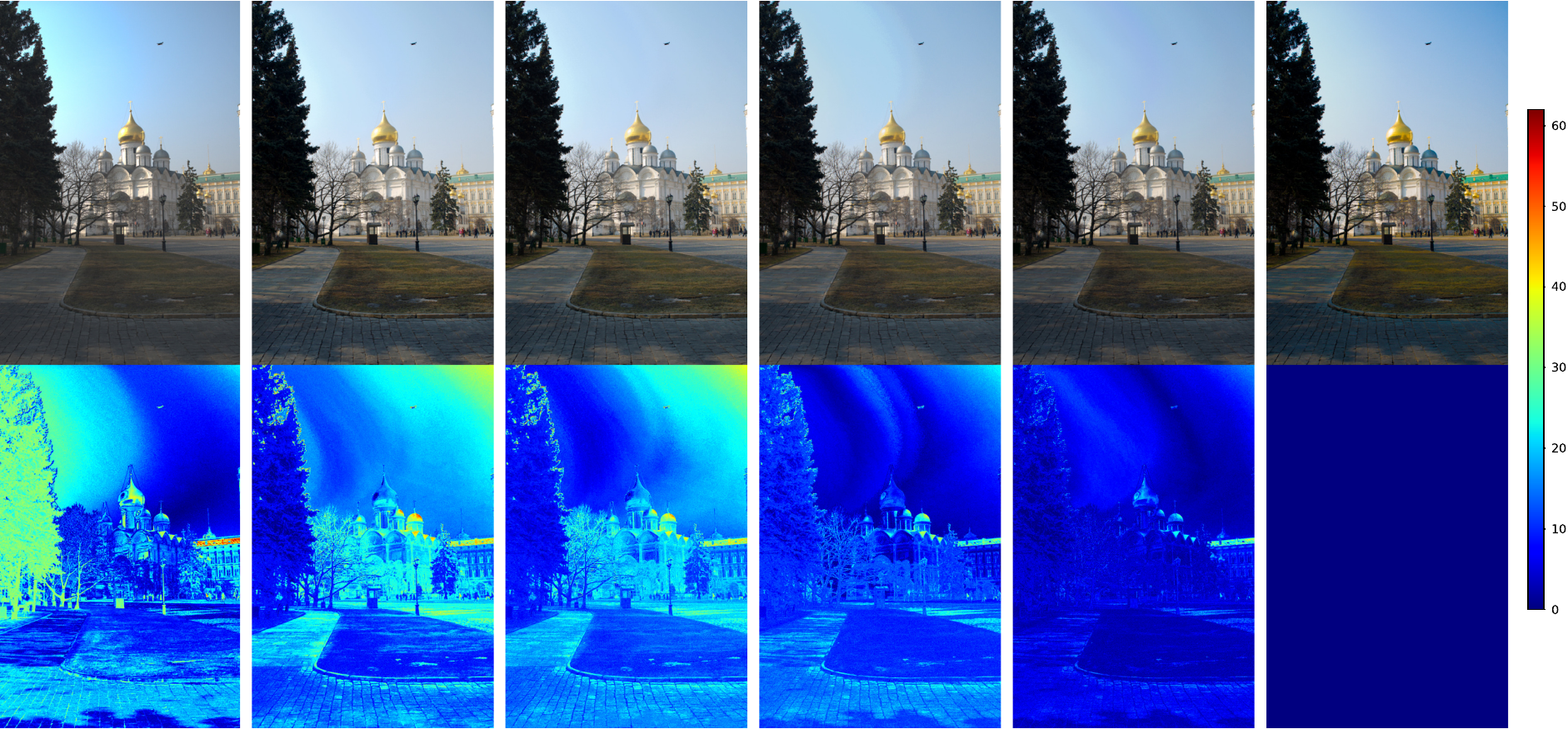}
      \label{fig:quali_fivek3}
    \end{subfigure} 
    \caption{Qualitative comparisons for photo retouch task on the FiveK dataset \cite{bychkovsky2011learning}.
    The error maps at the bottom of each picture present differences with ground truth.
    Each color on error map indicates the degree of error based on the corresponding color bars on the right.
    } 
    \label{fig:quali_fivek}
\end{figure*}

\begin{figure*}[tb]
    \centering
    \begin{subfigure}{0.85\linewidth}
      \includegraphics[width=\columnwidth]{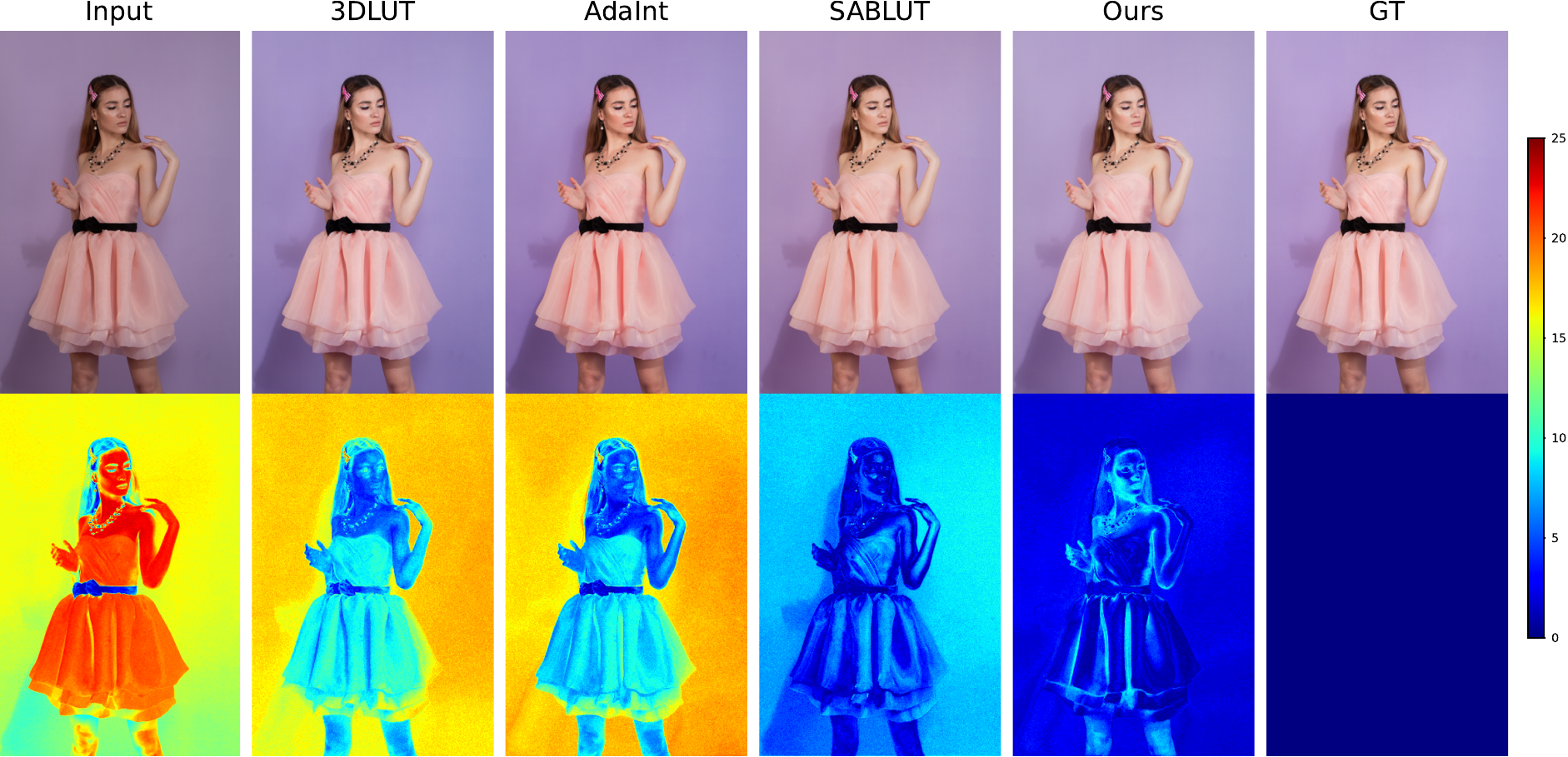}
     
      \label{fig:quali_ppr1}
    \end{subfigure}
    
    \begin{subfigure}{0.85\linewidth}
      \includegraphics[width=\columnwidth]{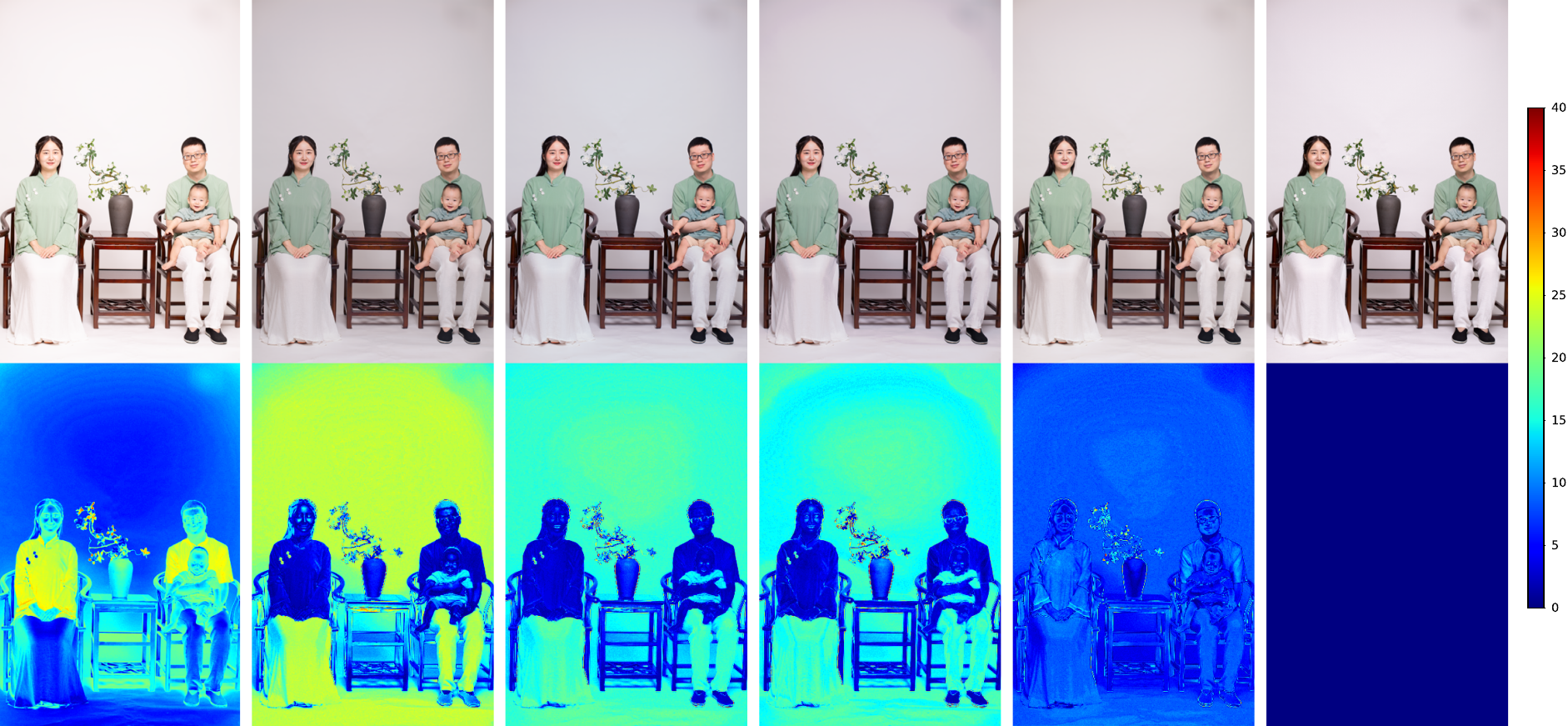}
      
      \label{fig:quali_ppr2}
    \end{subfigure}
    
    \begin{subfigure}{0.85\linewidth}
      \includegraphics[width=\columnwidth]{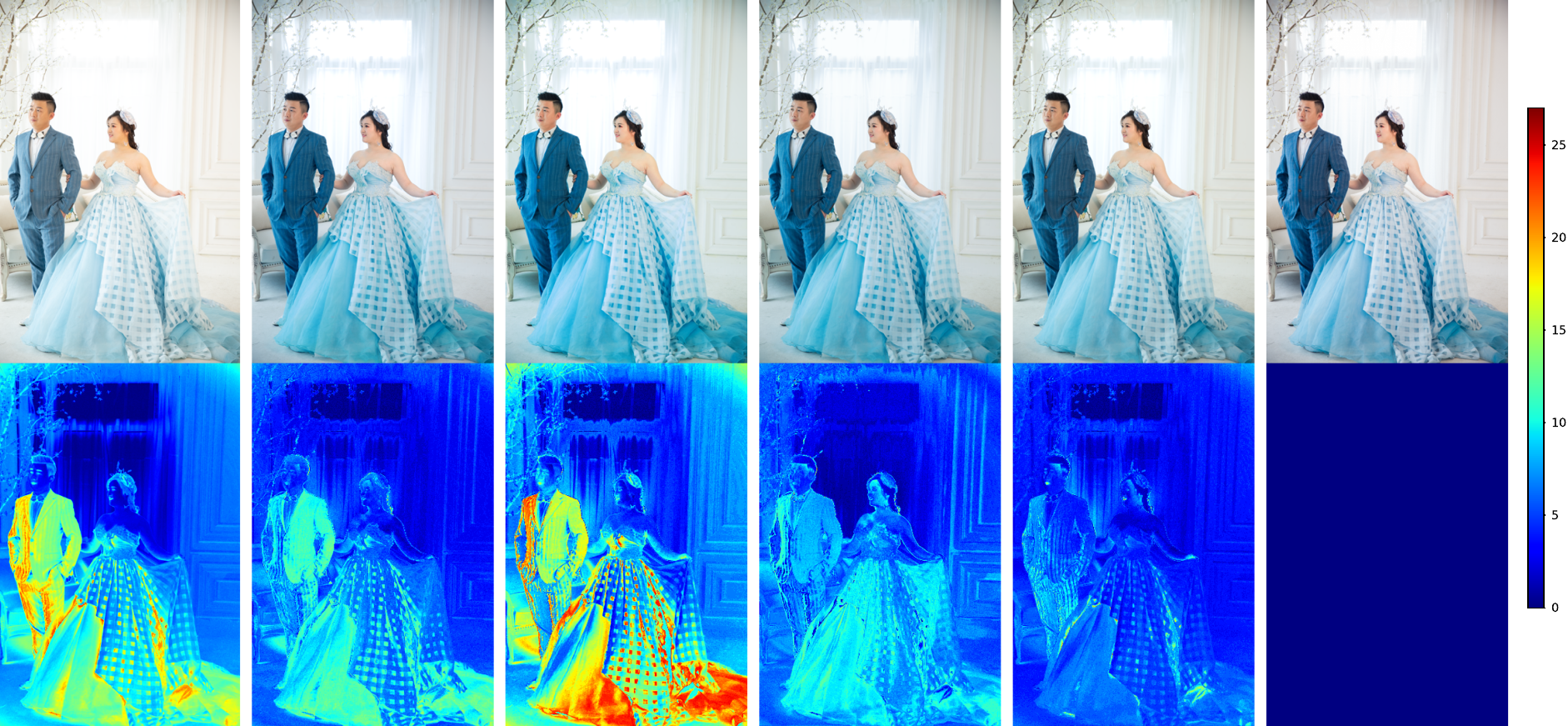}
      
      \label{fig:quali_ppr3}
    \end{subfigure} 
    \caption{Qualitative comparisons for photo retouch task on the PPR10K dataset \cite{liang2021ppr10k}.
    The error maps at the bottom of each picture present differences with ground truth.
    Each color on error map indicates the degree of error based on the corresponding color bars on the right.
    } 
    \label{fig:quali_ppr}
\end{figure*}

\begin{figure*}[tb]
    \centering
    \begin{subfigure}{\linewidth}
      \includegraphics[width=\columnwidth]{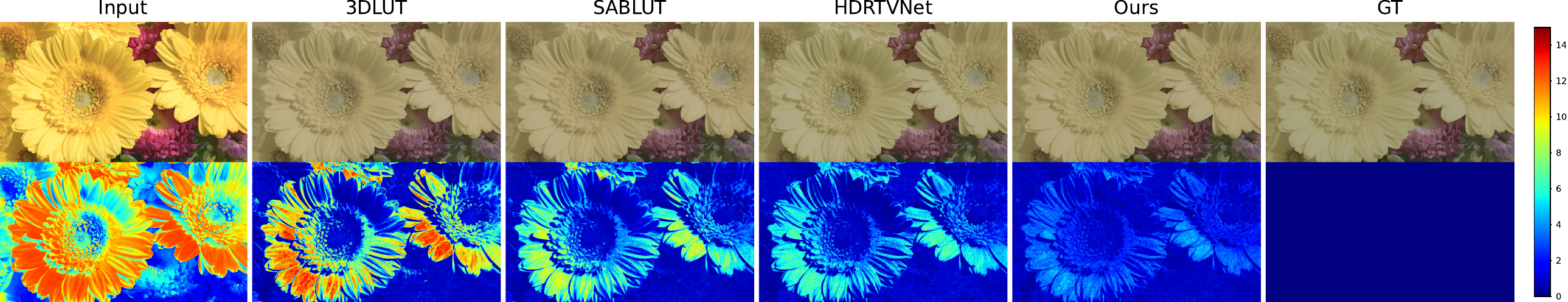}
      \label{fig:quali_hdrtv1}
    \end{subfigure}
    
    \begin{subfigure}{\linewidth}
      \includegraphics[width=\columnwidth]{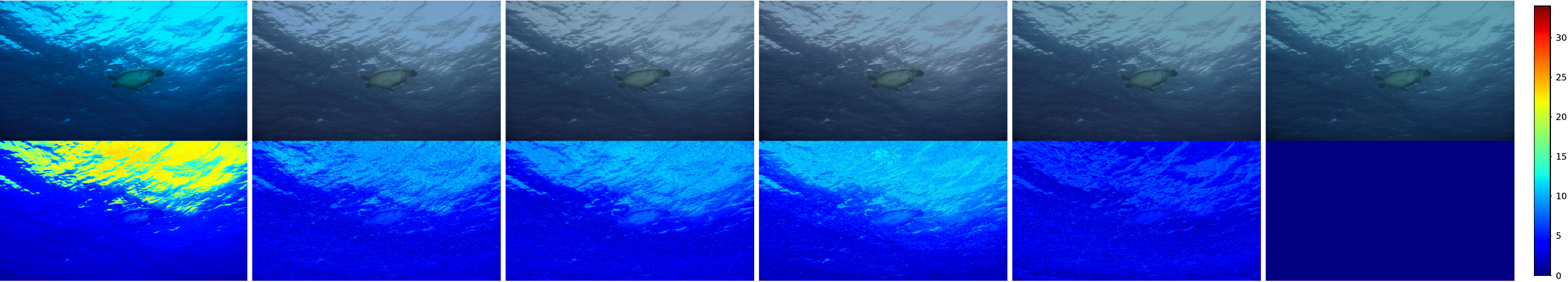}
      \label{fig:quali_hdrtv2}
    \end{subfigure}
    
    \begin{subfigure}{\linewidth}
      \includegraphics[width=\columnwidth]{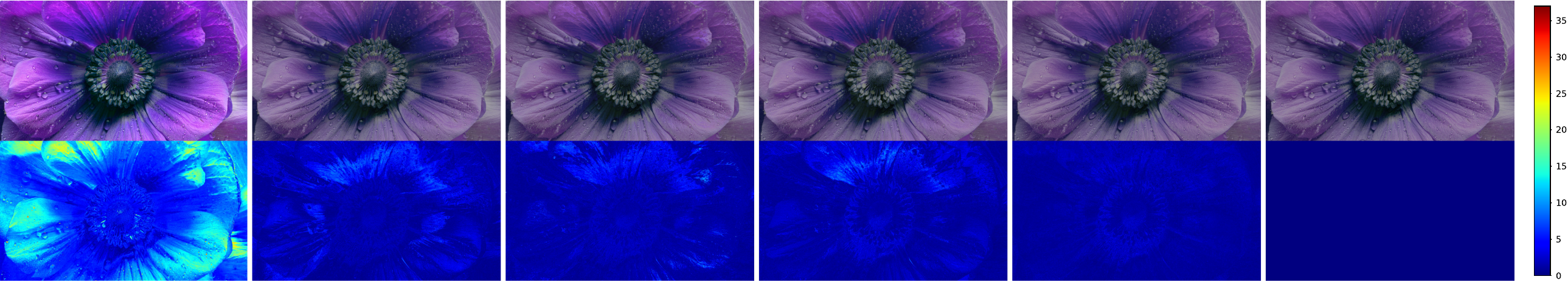}
      \label{fig:quali_hdrtv3}
    \end{subfigure} 
    \caption{Qualitative comparisons for SDRTV-to-HDRTV task on the HDRTV1K dataset \cite{chen2021new}.
    The error maps at the bottom of each picture present differences with ground truth.
    Each color on error map indicates the degree of error based on the corresponding color bars on the right.
    } 
    \label{fig:quali_hdrtv}
\end{figure*}

We provide additional qualitative results in \cref{fig:quali_fivek}, \cref{fig:quali_ppr} and \cref{fig:quali_hdrtv}.

\end{document}